\input harvmac
\def\x{ {x_{11}} }
\def\q{b}
\def\b{\td \q }
\def \inv {^{-1}}

\def\const {{\rm const}}
\def \s {\sigma}

\def \p {\phi}
\def \ha {\half}
\def \ov {\over}

\def \four{{\textstyle {1\ov 4}}}
\def \a {\alpha}
\def \lr {\lref }

\def\vp {\varphi}

\def \r {\rho}
\def\const {{\rm const}}

\def \g {\gamma}

\def \y {{ \tilde y}}

\def   \td {\tilde }

\def \lr { \lref }

\gdef \jnl#1, #2, #3, 1#4#5#6{ { #1~}{ #2} (1#4#5#6) #3}

\def \y {{\td y}}
\def \S {{\cal S}}


\def\del{\partial }

\def\const {{\rm const}}
\def \s {\sigma}

\def \ha {\half}
\def \ov {\over}

\def \four{{\textstyle {1\ov 4}}}
\def \a {\alpha}
\def \lr { \lref}

\def\vp {\varphi}

\def\const {{\rm const}}

\def\g {\gamma}

\def\y {{ \tilde y}}
 
\def   \td {\tilde }

\def \lr { \lref}
\def \bb {{\rm b}}
\def\r{r}
\gdef \jnl#1, #2, #3, 1#4#5#6{ { #1~}{ #2} (1#4#5#6) #3}

\lr \gibma {G.~W.~Gibbons and K.~Maeda,
``Black Holes And Membranes In Higher Dimensional
Theories With Dilaton Fields,''
Nucl.\ Phys.\ B {\bf 298}, 741 (1988).
}
\lr\silv{ A.~Adams and E.~Silverstein,
``Closed string tachyons, AdS/CFT, and large N QCD,''
hep-th/0103220.
}
\lr \EMP{R.~Emparan,
``Composite black holes in external fields,''
Nucl.\ Phys.\ B {\bf 490}, 365 (1997)
[hep-th/9610170].
}

\lr\gib{G.W.  Gibbons,
``Quantized Flux Tubes In Einstein-Maxwell Theory And
Noncompact Internal Spaces,''
in: {\it Fields and Geometry},
 Proceedings of the 22nd
Karpacz
Winter School of Theoretical Physics, ed.
 A. Jadczyk (World Scientific,
Singapore,  1986).}

\lr \tset { A.~A.~Tseytlin,
``Harmonic superpositions of M-branes,''
Nucl.\ Phys.\ B {\bf 475}, 149 (1996)
[hep-th/9604035].
}

\lr\GS{M.~Gutperle and A.~Strominger,
``Fluxbranes in string theory,''
hep-th/0104136.
 }

\lr \atick {J.~J.~Atick and E.~Witten,
``The Hagedorn Transition And The Number
Of Degrees Of Freedom
Of String Theory,''
Nucl.\ Phys.\ B {\bf 310}, 291 (1988).
}
\lr\ggrr{
M.~B.~Green,
``Wilson-Polyakov loops for critical strings
and superstrings
at finite temperature,''
Nucl.\ Phys.\ B {\bf 381}, 201 (1992).
  }
\lr\gal{ C.~Chen, D.~V.~Gal'tsov and S.~A.~Sharakin,
``Intersecting M-fluxbranes,''
Grav.\ Cosmol.\  {\bf 5}, 45 (1999)
[hep-th/9908132].
 }
\lr \NW {C.~R.~Nappi and E.~Witten,
``A WZW model based on a nonsemisimple group,''
Phys.\ Rev.\ Lett.\ {\bf 71}, 3751 (1993)
[hep-th/9310112].
}

\lr \tcqg {A.~A.~Tseytlin,
``Exact solutions of closed string theory,''
Class.\ Quant.\ Grav.\  {\bf 12}, 2365 (1995)
[hep-th/9505052].
}

\lr \niel{N.~K.~Nielsen and P.~Olesen,
``An Unstable Yang-Mills Field Mode,''
Nucl.\ Phys.\ B {\bf 144}, 376 (1978).
}

\lr \koun{ C.~Kounnas and B.~Rostand,
``Coordinate Dependent Compactifications
And Discrete Symmetries,''
Nucl.\ Phys.\ B {\bf 341}, 641 (1990).}
\lr \duff{ M.~J.~Duff, P.~S.~Howe, T.~Inami and K.~S.~Stelle,
``Superstrings In D = 10 From Supermembranes In D = 11,''
Phys.\ Lett.\ B {\bf 191}, 70 (1987).
 }
\lr \TP{A.~A.~Tseytlin,
``Type 0 strings and gauge theories,''
{\it Contributed to 14th International Workshop on High Energy Physics and
Quantum Field Theory } (QFTHEP 99), Moscow, June 1999.
(http://theory.sinp.msu.ru/~qfthep/$99_{-}$/Proceed99.html)
}
\lr \ort{ P.~Meessen and T.~Ortin,
``Type 0 T-duality and the tachyon coupling,''
hep-th/0103244.
}
\lr\russo{J.G. Russo, ``Free energy and critical temperature in eleven
dimensions", Nucl.\ Phys.\ B {\bf 602}, 109 (2001)
[hep-th/0101132].
}

\lr \suf { P.~M.~Saffin,
``Gravitating fluxbranes,''
gr-qc/0104014.
}

\lr\green{
J.~G.~Russo and A.~A.~Tseytlin,
``Green-Schwarz superstring action in a curved magnetic Ramond-Ramond
background,''
JHEP{\bf 9804}, 014 (1998)
[hep-th/9804076].
}

\lr \BG {O.~Bergman and M.~R.~Gaberdiel,
``Dualities of type 0 strings,''
JHEP{\bf 9907}, 022 (1999) [hep-th/9906055].
}

\lr\nucleation{ F.~Dowker, J.~P.~Gauntlett, G.~W.~Gibbons and G.~T.~Horowitz,
 ``Nucleation of $P$-Branes and Fundamental Strings,''
Phys.\ Rev.\ D {\bf 53}, 7115 (1996)  [hep-th/9512154].
}

\lr\magnetic{
J.~G.~Russo and A.~A.~Tseytlin,
``Magnetic flux tube models in superstring theory,''
Nucl.\ Phys.\ B {\bf 461}, 131 (1996)
[hep-th/9508068].}

\lr\heterotic{
J.~G.~Russo and A.~A.~Tseytlin,
``Heterotic strings in uniform magnetic field,''
Nucl.\ Phys.\ B {\bf 454}, 164 (1995)
[hep-th/9506071].}

\lr\exactly{
J.~G.~Russo and A.~A.~Tseytlin,
``Exactly solvable string models of curved space-time backgrounds,''
Nucl.\ Phys.\ B {\bf 449}, 91 (1995)
[hep-th/9502038].}

\lr\constant{
J.~G.~Russo and A.~A.~Tseytlin,
``Constant magnetic field in closed string theory: An Exactly solvable model,''
Nucl.\ Phys.\ B {\bf 448}, 293 (1995)
[hep-th/9411099].
G.~T.~Horowitz and A.~A.~Tseytlin,
``A New class of exact solutions in string theory,''
Phys.\ Rev.\ D {\bf 51}, 2896 (1995)
[hep-th/9409021].
}
\lr \BD { J.~D.~Blum and K.~R.~Dienes,
``Strong/weak coupling duality relations for non-supersymmetric
string  theories,''
Nucl.\ Phys.\ B {\bf 516}, 83 (1998)
[hep-th/9707160].
 M.~Dine, P.~Huet and N.~Seiberg,
``Large And Small Radius In String Theory,''
Nucl.\ Phys.\ B {\bf 322}, 301 (1989).
  }

\lr \PT { G.~Papadopoulos and P.~K.~Townsend,
``Intersecting M-branes,''
Phys.\ Lett.\ B {\bf 380}, 273 (1996)
[hep-th/9603087].
}
\lr \gaun {F.~Dowker, J.~P.~Gauntlett, G.~W.~Gibbons and
G.~T.~Horowitz,
``The Decay of magnetic fields in Kaluza-Klein theory,''
Phys.\ Rev.\ D {\bf 52}, 6929 (1995)
[hep-th/9507143].
F.~Dowker, J.~P.~Gauntlett, S.~B.~Giddings and G.~T.~Horowitz,
``On pair creation of extremal black holes and Kaluza-Klein
monopoles,''
Phys.\ Rev.\ D {\bf 50}, 2662 (1994)
[hep-th/9312172].
F.~Dowker, J.~P.~Gauntlett, D.~A.~Kastor and J.~Traschen,
``Pair creation of dilaton black holes,''
Phys.\ Rev.\ D {\bf 49}, 2909 (1994)
[hep-th/9309075].
}
\lr\closed{A.~A.~Tseytlin,
``Closed superstrings in magnetic field:
 instabilities and supersymmetry breaking",
Nucl.\ Phys.\ Proc.\ Suppl.\  {\bf 49}, 338 (1996)
[hep-th/9510041].
  }
\lr\tse{A.~A.~Tseytlin,
``Melvin solution in string theory,''
Phys.\ Lett.\ B {\bf 346}, 55 (1995)
[hep-th/9411198].
}
\lr\imam{Y.~Imamura,
``Branes in type 0 / type II duality,''
Prog.\ Theor.\ Phys.\  {\bf 102}, 859 (1999)
[hep-th/9906090].
B.~Craps and F.~Roose,
``NS fivebranes in type 0 string theory,''
JHEP {\bf 9910}, 007 (1999)
[hep-th/9906179].
}
\lr \typ{L.~J.~Dixon and J.~A.~Harvey,
``String Theories In Ten-Dimensions Without Space-Time
Supersymmetry,''
Nucl.\ Phys.\ B {\bf 274}, 93 (1986).
N.~Seiberg and E.~Witten,
``Spin Structures In String Theory,''
Nucl.\ Phys.\ B {\bf 276}, 272 (1986).
C. Thorn, unpublished.
}

\lr\prt{ G.~Papadopoulos, J.~G.~Russo and A.~A.~Tseytlin,
``Curved branes from string dualities,''
Class.\ Quant.\ Grav.\ {\bf 17}, 1713 (2000),
hep-th/9911253.}

\lr \CG {M.~S.~Costa and M.~Gutperle,
``The Kaluza-Klein Melvin solution in M-theory,''
hep-th/0012072.}

\lr \bst{ E.~Bergshoeff, E.~Sezgin and P.~K.~Townsend,
``Properties Of The Eleven-Dimensional Super Membrane Theory,''
Annals Phys.\ {\bf 185}, 330 (1988).}
\lr \melvv{ M.A. Melvin,  Phys.\ Lett.\ B {\bf 8}, 65 (1964).}
\lr \rohm {R.~Rohm,
``Spontaneous Supersymmetry Breaking In Supersymmetric String
Theories,''
Nucl.\ Phys.\ B {\bf 237}, 553 (1984).
}
\lr \IGO {I.~R.~Klebanov,
``Tachyon stabilization in the AdS/CFT correspondence,''
Phys.\ Lett.\ B {\bf 466}, 166 (1999)
[hep-th/9906220].
}
\lr \KT { I.~R.~Klebanov and A.~A.~Tseytlin,
``D-branes and dual gauge theories in type 0 strings,''
Nucl.\ Phys.\ B {\bf 546}, 155 (1999)
[hep-th/9811035].
}

\lr \KTt { I.~R.~Klebanov and A.~A.~Tseytlin,
``A non-supersymmetric large N CFT from type 0 string theory,''
JHEP {\bf 9903}, 015 (1999)
[hep-th/9901101].
}

\lr\ttt {A.~A.~Tseytlin,
``Sigma model approach to string theory effective actions with
tachyons,''
 J. Math. Phys. {\bf 42}, 2854 (2001) [hep-th/0011033].
 }

\lr \HT { C.~M.~Hull and P.~K.~Townsend,
``Unity of superstring dualities,''
Nucl.\ Phys.\ B {\bf 438}, 109 (1995)
[hep-th/9410167].}

\lr\callan{
A.~Abouelsaood, C.~G.~Callan, C.~R.~Nappi and S.~A.~Yost,
``Open Strings In Background Gauge Fields,''
Nucl.\ Phys.\ B {\bf 280}, 599 (1987).}

\lr\sw{N.~Seiberg and E.~Witten,
``String theory and noncommutative geometry,''
JHEP{\bf 9909}, 032 (1999),
hep-th/9908142.}

\lr \JS {J.~H.~Schwarz,
``An SL(2,Z) multiplet of type IIB superstrings,''
Phys.\ Lett.\ B {\bf 360}, 13 (1995)
[hep-th/9508143].}
\baselineskip8pt
\Title{\vbox
{\baselineskip 6pt{\hbox{OHSTPY-HEP-T-01-009  }}{\hbox
{NSF-ITP-01-43}}{\hbox
{}}{\hbox{hep-th/0104238}} {\hbox{
   }}} }
{\vbox{\centerline {Magnetic backgrounds and tachyonic
instabilities}
\vskip4pt
\centerline{ in closed superstring theory
and M-theory   }
 }}
\vskip -20 true pt
\centerline  {  {J.G. Russo$^{a,b}$
\footnote {$^*$} {e-mail address: russo@df.uba.ar}
and
A.A. Tseytlin$^c$
\footnote{$^{\star}$}{\baselineskip8pt
e-mail address: tseytlin.1@osu.edu}
\footnote{$^{\dagger}$}{\baselineskip8pt
Also at Imperial College, London and  Lebedev  Physics
Institute, Moscow.}
 }}
 \medskip \smallskip
 \centerline {\it {}$^a$ Departamento de F\'\i sica, Universidad de Buenos
Aires, }
\smallskip
\centerline {\it  Ciudad Universitaria, Pab. I, 1428 Buenos Aires, 
and Conicet}

\medskip

 \centerline {\it {}$^b$ The Abdus Salam International Center for Theoretical
Physics, }
\smallskip
\centerline {\it  Strada Costiera 11, 34014 Trieste, Italy.}

\medskip

\centerline {\it  {}$^c$ Smith Laboratory, Ohio State University, }
\smallskip

\centerline {\it   Columbus OH 43210-1106,
USA }
\bigskip
\centerline {\bf Abstract}
\medskip
\baselineskip10pt
\noindent
Models of closed superstrings
in certain curved  NS-NS magnetic flux backgrounds are exactly solvable
in terms of free fields.
They  interpolate
between free superstring theories
with periodic and antiperiodic boundary conditions
for fermions around some compact direction, and,
in particular, between type 0 and type II string theories.
Using  ``9--11" flip, this interpolation can be  extended to M-theory
and provides an interesting
 setting for  study of  tachyon problem in closed string theory.
Starting  with a general
 2-parameter family of such Melvin-type models,
we  present several  new
 magnetic flux backgrounds in 10-d string theory and 11-d M-theory
 and discuss their tachyonic instabilities.
In particular, we suggest
 a description of type 0B theory in terms of  M-theory
 in curved
  magnetic flux background, which supports its conjectured
 $SL(2,Z)$ symmetry,
  and in which  the type 0 tachyon   appears to  correspond
 to a state in  $d=11$ supergravity multiplet.
In another ``T-dual" description, the tachyon
 is related to a winding membrane state.
\medskip

\Date {April 2001}

\noblackbox
\baselineskip 16pt plus 2pt minus 2pt


\vfill\eject
\def \N {{\hat N}}

\def \y {{\td y}}
\def \S {{\cal S}}

\def\p{\partial }

\newsec{Introduction}
One  of the  fundamental problems in string theory is
 vacuum stability. Non-supersymmetric vacua  are
often unstable,   containing tachyons. Recently, there was a progress
in  understanding the issue of open string tachyon condensation.
The closed string tachyon problem is harder, and to
  try to address it,
type 0 string theory \typ\  is a natural  model to study.
As was noted in \ttt, in contrast to what happens in the open
string case, the type 0 string partition function
in tachyon background
 does not seem to
contain  a  tree-level
potential for the tachyon, i.e. a stable vacuum cannot be
determined just   from analysis of low-energy tree-level  effective theory
in flat background.

There have been   two proposals on
tachyon stabilization in type 0 theory -- one within string
 perturbation theory but in curved background,   and
another based on  non-perturbative M-theory formulation:

\noindent
(i)   it was suggested in  \refs{\KT,\KTt}
that type 0B string
in $AdS_5 \times S^5$ +  5-form R-R   background
  becomes stable for small enough  scale  of the
   space;\foot{This suggestion is motivated by the fact that
   R-R flux shifts
    the tachyon (mass)$^2$  in positive direction.
   To prove  this conjecture
   one needs to demonstrate  that $\a'$-corrections
   do not spoil  this effect, i.e. to  solve
    the corresponding R-R sigma model.
   Using a generalization of AdS/CFT correspondence,
   for $\a' R \sim 1$  one may expect  to
   get some input from the dual  large N  gauge theory
   description
 which is   weakly coupled in that regime \IGO.
    For a recent interesting discussion  of    issues
 related to closed string tachyons
     in orbifold theories in the  context of AdS/CFT correspondence 
(and a suggestion that type 0  theory may still remain unstable
at small radius  due
to a new tachyon appearing in the twisted sector)  see 
    \silv.}

    \noindent
 (ii)  it was conjectured  in    \BG\
 that the tachyon of type 0A string in flat space
  gets  $m^2>0  $ at strong coupling (where type 0A string
    becomes dual to
  11-d M-theory on  large circle with antiperiodic fermions).
  Ref.  \CG\   combined this second proposal
  with the observation
 that Kaluza-Klein  Melvin \melvv\ 
  magnetic background  \refs{\gibma,\gib,\gaun,\nucleation}
  in type II string theory  and in  11-d  M-theory
   may be used to
  interpolate \refs{\magnetic,\green} between
  periodic and antiperiodic boundary conditions  for
  space-time fermions.
  This suggests  a dynamical mechanism of type 0 string  tachyon
  stabilization -- it  disappears at strong  magnetic R-R field
   \refs{\CG,\GS}.

Our aim  will be  to extend the discussion  in \refs{\green,\CG}
to the case of more general class of  Melvin-type backgrounds
with two independent magnetic parameters $\q$ and $\b$.
It  includes K-K Melvin ($\q\not=0, \b=0)$
 and dilatonic Melvin ($\q = \b)$ \refs{\gibma,\tse}
 solutions
 as special cases and  is covariant under T-duality.
The  bosonic \exactly \ and superstring
 \refs{\magnetic,\closed } \  models
in the corresponding NS-NS backgrounds  are exactly solvable, with perturbative
spectrum and torus partition
function known explicitly (see \tcqg\ for a review).

Here we will lift these type IIA magnetic  solutions
to 11 dimensions,  getting a $\b\not=0$
generalization of the flat \refs{\nucleation,\gaun}
$d=11$ background discussed  in \refs{\green,\CG}.
 Dimensional reduction along different directions (or U-dualities
 directly in $d=10$) lead to a number of $d=10$ supergravity
 backgrounds with
 R-R magnetic fluxes, generalizing  to $\b\not=0$
 the R-R magnetic  flux 7-brane \refs{\green,\CG}  of type IIA theory.
 We shall  investigate instabilities
 of these backgrounds.

 Our motivation  is to get a  more detailed understanding
 of magnetic interpolations between stable and unstable
 theories, combining previous perturbative results
 \refs{\magnetic} about spectra of NS-NS models  with
 non-perturbative  $d=11$   extensions  based,
  as in  \refs{\BG,\CG},
  on  conjectured  applicability of the  ``9-11"  flip.
 One  general  aim  is to  establish  precise relations
  between  non-supersymmetric
   backgrounds in  type II superstring
   theory  and  unstable backgrounds in  type 0 theory.
   Another is to  use  interpolating magnetic backgrounds
   to connect   D-brane solutions
    in the two theories,
     and, hopefully,
   shed more light on  non-supersymmetric gauge theories
   dual to them. In the process, one may  also
    to learn  more
   about  various non-supersymmetric compactifications of M-theory.

\bigskip


Before describing the  content of the paper let us recall
some basic points about the relation between
 type 0 and type II   theories.
First,  there is a  perturbative  relation in flat space:
weakly coupled
type 0 string theory  may be  interpreted as an orbifold
of type II theory \refs{\BD}.  Indeed, type II  theory
compactified on a
circle with {\it anti}periodic
 boundary conditions for space-time fermions
 is closely related
 to type 0  theory compactified on a circle \refs{\rohm,\atick}.
More  precisely, the two theories
are limits of the  interpolating  ``9-dimensional"
theory \BD\ --  $\Sigma_R$  orbifold of type II theory.
 $\Sigma_R$ stands for $  (S^1)_R/[(-1)^{F_{s} }\times {\cal S}]$,
 where
 $F_s$ is space-time fermion number and
 $\S$ is half-shift along the circle ($X_9 \to X_9 + \pi R$).\foot{The
 role of ${\cal S}$ is to mix $(-1)^{F_{s} }$  orbifold
 with compactification on the circle,
  allowing for the  interpolation
  \BD. Invariant states under ${\cal S}$  have  even momentum quantum
  numbers $m$ and integer winding numbers $w$, while
  twisted sector states  have  odd $m$ and half-integer $w$.
  The action  of $\cal S$ is irrelevant for $R\to \infty$, but
  is crucial for reproducing type 0 spectrum for $R\to 0$.
  The existence of this  9-d
   interpolating
   orbifold theory allows one to establish
   relations between  D-branes in type II and
   type 0 theories \refs{\KTt,\imam}.}
 Type IIA on $\Sigma_{R\to \infty}$  is
 type IIA theory  on an infinite circle
  and  type IIA on $\Sigma_{R \to 0}$ is
  type 0A theory  on $(S^1)_{R\to 0}$ (or T-dual  type 0B theory
 on  $(S^1)_{ R\to \infty}$). Thus
  type IIA on $\Sigma$ interpolates between  type IIA
 and type 0B {\it ten}-dimensional theories \BD.\foot{
 In general, there are 4 different perturbative
 interpolating
 9-d theories \BD: (a) type IIA on $\Sigma_{R}$
 (relating   d=10  type IIA and type  0B);
 (b) type IIB on $\Sigma_{R}$
 (relating   d=10  type IIB and type  0A);
 (c) type IIA on $(S^1)_{R}$ = type IIB on $(S^1)_{\tilde R}$,
 $\tilde R = \a'/R$
 (relating   d=10  type IIA and type  IIB);
 (d) type 0A on $(S^1)_{R}$ = type 0B on $(S^1)_{\tilde R}$,
 $\tilde R = \a'R^{-1}$
 (relating   d=10  type 0A and type  0B).
 These theories are not equivalent -- they
 share only  respective boundary points.
 The   $\Sigma_{R}$     orbifold of type IIA(B) theory
 is, at the same time, equivalent (T-dual)
 to the  $S^1/[(-1)^{F_{\rm r}}\times {\cal S}]$
 orbifold of type 0B(A) theory
 ($F_{\rm r}$ is right-moving world-sheet fermion number)
  \BG.
  These interpolating orbifold theories have, in contrast to type 0
  theory,  massive fermions in their spectrum.}

Using that type II superstring
in the  $d=10$ K-K    Melvin  NS-NS background
$ds^2=-dt^2 + dx_s^2+dx_9^2+dr^2+ r^2  (d\varphi +\q \ dx_9)^2$
(with $x_9 \equiv x_9 + 2 \pi R$)
is,   for $\q R=1$,  equivalent
  to type II theory on $R^9 \times (S^1)_R$
  with antiperiodic boundary
 conditions for the  space-time fermions
 along the $x_9$ circle  \magnetic,\foot{This
  model at $\q R=1$ becomes
 essentially  the same as  a
 (non-compact 2-plane)   limit of  the
 twisted 3-torus model of \rohm\ or a Wick rotation of the
 finite temperature superstring  theory \refs{\atick,\ggrr}.}
 one may   notice  \CG\ that
 it is  equivalent  to  the  above orbifold  of
 type II on $\Sigma_{R'}$   with  the
  radius $R'= 2 R$ (extra  2 is related to the  shift
   ${\cal S}$).\foot{In the Melvin model  with
   $\q R=1$ the shift  $x_9 \to x_9 + 2\pi R$
   is equivalent to $2\pi$ rotation in the 2-plane under which
   space-time fermions  change sign. In the orbifold,
   the corresponding shift  $\cal S$ is  by $ \pi R'$.}
    Therefore,  this  NS-NS Melvin model
 describes, in the limit $R\to 0$,
  weakly coupled  type 0 string on  $R^{1,8}\times
 (S^1)_{R\to 0}$ (or type 0B on $R^{1,8}\times
 (S^1)_{R\to \infty}$).\foot{Combined with the results
 of \magnetic,
 this observation  allows one to describe type 0 theory
 in
 $R^{1,8}\times (S^1)_{R\to \infty}$ in  light-cone Green-Schwarz formulation,
 where the action is quadratic  in (standard, periodic
 in $\sigma$) fermions,
 but fermions  are not decoupled from the radial
 coordinate of $S^1$ -- their   covariant derivative has
  flat but topologically non-trivial
 connection $\sim  \q \del x_9$. Redefining the
 fermions  to eliminate this connection makes them
 antiperiodic in $\sigma$ in the odd winding sector
   \refs{\magnetic,\closed}.}
 To summarize, type II orbifold on $\Sigma_{2R}$  is the same as
 type II theory in Melvin background at $\q R=1$ and
 thus the latter model  interpolates between type II
 and type 0 theories in  infinite $d=10$ space.

 This can be extended to a general claim \CG\
 that type IIA(B) theory
 in the Kaluza-Klein   NS-NS  Melvin background
 with parameters $(\q , R)$
  is equivalent
 (has the same perturbative spectrum and thus
  same torus partition
 function)  to the   orbifold of
 type IIA(B)  on $\Sigma_{R'}$
  in this  NS-NS  Melvin background with
  parameters
$(\q'= \q - R^{-1},\ R'= 2R)$.

 Returning back to the case of flat background, the above
 compactification of type IIA theory on $\Sigma_{R_9}$  corresponds
  to M-theory
 on $(S^1)_{R'_{11}} \times (S^1)_{R_{9}}/[(-1)^{F_{s}
 }\times {\cal S}]$, where
 $R_9$ and $R'_{11}$
 are  taken to be small.
  Making the 9-11 flip\foot{This  means exchanging the roles
  of the two circles,  assuming that there is indeed an interpolating
 coordinate-invariant M-theory in 11 dimensions.
 Applicability of this flip
 remains  of course a conjecture.}
 one  may  then conjecture \BG\
 that one  gets an equivalent
 description of this  theory
 as an ordinary  $S^1$ (i.e. periodic fermions)
compactification of the $d=10$ theory obtained from M-theory on
$\Sigma_{R'_{11}}$.  Then
type 0A  theory may be interpreted as
   M-theory on $
   \Sigma_{R'_{11}}$, or, essentially,
   as M-theory on  $S^1$ with  radius
   ${ 1 \ov 2}  R'_{11}$ with periodic/antiperiodic
   boundary conditions for bosons/fermions.\foot{
 The factor of $1/2$ is
   again due to the half-shift $\S$. It  suggests that 
    the same  factor should be  present 
   in the relation between string coupling constants.
   However,  it is not clear how  to decide this 
   unambiguously since the relation to perturbative 
    type 0 theory based on 9-11 
   flip applies only in the limit of zero radius, i.e. at 
   zero coupling.}
Type 0B theory in $R^{1,9}$
is then  M-theory on $T^2/[(-1)^{F_{s}
 }\times {\cal S}]$ in the limit of zero volume of the
  2-torus \BG ,  and
it was argued in \BG\ that,   like type IIB theory,  this theory
should be $SL(2,Z)$ symmetric.


Lifting the above  NS-NS   Melvin metric to 11 dimensions,
replacing $x_9 \leftrightarrow x_{11}$ and reducing it down to
10 dimensions  along $x_{11}$  gives a   non-supersymmetric
 R-R  Melvin   7-brane in
type IIA theory \refs{\nucleation,\green}.
The dilaton  $e^\phi = g_s ( 1 + \q^2 r^2)^{3/4} $
grows away from the core, i.e. for large $r$
 the theory becomes  effectively
11-dimensional \green.

Combining the above observations, it was suggested in \CG\
that type IIA  and type 0A theories  in such $d=10$
 R-R Melvin backgrounds\foot{Any  bosonic solution
 of type II supergravity can be embedded into type 0 theory
 provided the fields of the twisted sector (tachyon and
 second set of R-R bosons) are set equal to zero \refs{\KT}.}
   are non-perturbatively
   dual to each other, with parameters  related by
 $ \q_0= \q- R^{-1}_{11}, \ g_{s0} 
= { R'_{11}\ov   2 \sqrt { \a'} }
 ={ R_{11}\ov   \sqrt { \a'} }$.
 For example,  for $\q R_{11}=1$ the $d=11$  Melvin theory  has  
  antiperiodic fermions on the   circle $R_{11}$, while, according 
  to \BG, type 0 theory  is M-theory with antiperiodic 
  fermions on circle $\ha R'_{11}$.
 That means, in particular, that starting with type 0 theory
 and increasing  the  value of the R-R magnetic
 parameter $\q_0$, the tachyon should disappear at strong
 enough magnetic field \CG\  when the theory
  becomes strongly coupled --  its
 weakly-coupled description  is in terms of
  stable weakly coupled type II theory
 in R-R Melvin background with small $\q $.

\bigskip

This paper is organized as follows.
In section 2 we review the 2-parameter $(\q,\b)$
  NS-NS Melvin
type II superstring model of \magnetic  . We describe the
periodicity of the mass spectrum
in magnetic field parameters, and show
how at special values of the
 parameters the spectrum reduce
 to that of the
free superstring theory with fermions antiperiodic in spatial direction
(reproducing, in particular,  the perturbative type 0 spectrum
in the $R\to 0$ limit).
We also  discuss tachyonic
instabilities, and  explain how they can be investigated
using the effective field theory.

In section 3 we consider the  U-dual   2-parameter R-R Melvin
magnetic
background, and find the corresponding  11-dimensional solution.
This 11-d background is no longer locally flat,
 in contrast to the  K-K Melvin case $(\q\not=0,\b=0)$
considered in  \refs{\green,\CG}.

In section 4 we discuss two alternative representations of
type 0A/0B string theories in terms
of M-theory compactifications in the magnetic backgrounds, and study
the appearance of  instabilities as   magnetic field is varied.
The  representation of type 0B theory in terms of M-theory in
curved
$(\q=0,\b\not=0)$  magnetic background
makes manifest  its   conjectured  $SL(2,Z)$  symmetry.
We also discuss the role
of  winding membrane states in reproducing  type 0 tachyon
from M-theory perspective, and  comment on some implications
 for M-theory  at finite temperature.

In section 5 we  consider the equation for the tachyon
of weakly-coupled  type 0A string theory
in the  R-R Melvin background  and demonstrate that,  as in
 \refs{\KT,\KTt}, the R-R flux shifts
 the value of tachyon (mass)$^2$
 in positive direction. As we explain,
 this fact may be  considered as an additional argument
 in favor of the    the conjecture
 that the type 0A  tachyon  should disappear
 at sufficiently strong  magnetic R-R field.

In section 6  we present  a general class of  $d=11$
 magnetic flux brane solutions with several
independent  field parameters,
 generalizing  the background  of section 3.
Solutions representing magnetic fluxbranes
  superposed with ordinary  D-brane backgrounds
   may  dynamically relate stable and  unstable
   string theory configurations with R-R charges,
 in particular, type IIA branes and  type 0A branes.
 Such  solutions  that can be obtained from  $d=11$
 flux branes
 superposed with charged M2-branes  and M5-branes,
are constructed in section 7.

In all of these cases, the magnetic fluxes
are concentrated near $r=0$ and vanish at $r=\infty $.
In  the Appendix, we  describe  a different  solution
with a {\it uniform} R-R magnetic field,
 representing  rotating space-time  dual to the NS-NS
 background  in \constant.

\newsec{String models with NS-NS magnetic flux backgrounds
}

In this section  we shall  summarize the results
of \magnetic\ for the type II superstring
   models  which  describe
  static  magnetic flux tube  backgrounds.
The  bosonic part of the corresponding conformal
 $\sigma $ model has the  Lagrangian
 $$
 L= \del_+ x_i \del_- x_i  + \del_+ \r  \del_- \r
 +   { \r^2 \ov  1 +\b ^2 \r ^2 }
[ \del_+ \vp  +  (\q  +  \b )    \del_+ x_9 ]
 \  [  \del_- \vp + (\q  -\b )  \del_- x_9 ]   $$
\eqn\lagg{
  +  \  \del_+ x_9  \del_- x_9  +
  {\cal R} [\phi_0 -  \ha \ln  (1+\b ^2 \r ^2 )] \  .}
Here $x_i=(t,x_s), \ s=1,...,6$  are  the free string coordinates,
 $x_9$ is a periodic coordinate of radius $R=R_9$ and  $\vp $
 is the  angular  coordinate with period
 $2 \pi$ ($ {\cal R} $ is  proportional to a background world-sheet
 curvature).
 The constants $\q$ and $\b$ are the magnetic field
 parameters.
Explicitly, the   curved  10-d
background geometry
has the following  string-frame metric, NS-NS 2-form and the
dilaton
\eqn\melv{
ds^2=-dt^2 + dx_s^2+dx_9^2+dr^2+ {r^2\over 1 +\b ^2 r^2}  [d\varphi
+(\q +\b  ) dx_9] [d\varphi +
(\q -\b  ) dx_9]\ ,
}
\eqn\meel{B_2 ={\b  r^2\over 1+\b  ^2 r^2} d \vp \wedge dx_9 \ , \ \ \ \ \ \ \ \
\ \ \ \
e^{2(\phi -\phi_0)}={1\over 1+\b ^2 r^2}\ ,
}
i.e.  may be interpreted as a 6-brane background
in $d=10$ type II string theory.
The magnetic Melvin-type interpretation becomes
apparent  upon dimensional reduction in the $x_9$-direction.
The resulting solution of $d=9$ supergravity
\eqn\backg
{ ds^2_9=  -dt^2 + dx_s^2   + d\r^2 +   {\r^2 f^{-1}  \tilde f^{-1}}
d\vp^2  \ ,\
}
\eqn\baag{ {\cal A}_\vp=  \q      \r^2 f\inv  \ , \ \ \ \ \ \ \
{\cal B}_\vp=  -\b    \r^2 \td f\inv  \ , \ }
\eqn\deq{
e^{2(\phi-\phi_0)}=\td f^{-1} \ ,\ \   e^{2\s} =
 f  \td f \inv
 \ , \ \ \ \ \ \ \ \ \ \ \
  f  \equiv 1 +  \q ^2 \r^2  \  ,\ \ \
\tilde f  \equiv  1 + \b  ^2 \r^2 \ , }
 describes  a magnetic flux tube universe with
 the magnetic  vector ${\cal A}$ (coming from the metric)
  and  the axial-vector $ {\cal B}$ (coming from the 2-form)
and the non-constant
dilaton $\phi$ and K-K  scalar $\s$. Note the
 $\q \leftrightarrow \b$
 (T-duality)  covariance of this background.

The special case of $\b=0$ corresponds to a
 locally flat 10-d metric
(thus the model \lagg\ with $\b=0$ can  be
formally  obtained from the free string model  by the shift
$\varphi \to \varphi +\q  x_9$)
\eqn\fff{
ds^2=-dt^2 + dx_s^2+dx_9^2+dr^2+
r^2  (d\varphi +\q\  dx_9)^2  \ , }
but the  9-d geometry \backg\ is still curved.
 In the notation of \gibma , the model with
 $\b =0 $ and  arbitrary $\q $ corresponds
   to the ``$a=\sqrt{3} $" (Kaluza-Klein) Melvin model
   (i.e. Melvin solution in $d=4$ Einstein  gravity
coupled to a Kaluza-Klein $U(1)$ gauge field),
whereas the one  with  $\b =\q $  is the  ``$a=1$" (dilatonic)
 Melvin model. The model  with $\q=0, \b\not=0$
 is T-dual to the $\b=0, \q\not=0$  model  and, as follows from \melv,
  has the following curved
 $d=10$ metric
\eqn\beb
{ ds^2_{10}=  -dt^2 + dx_s^2   + d\r^2 + \td f^{-1} (dx_9^2 +  {\r^2 }
d\vp^2)  \ .\
}
The $(R, \b ,\q )$ model
is  T-dual ($x_9 \to \td x_9$)
 to $(\td R, \q , \b )$ model, with
\eqn\pee{
\td R={\a'\over R}\ . }
Note that  the $\q =\b$
  ($a=1$ dilatonic Melvin)   model with $R=\tilde R$ is   the   `self-dual'
point.
   In view of this T-duality relation,
    one may say that
 string models with $\b  >\q  $ are effectively
equivalent  to  models with $\b <\q $.
 For fixed $\q $ inequivalent  models  are parametrized
 by
 $\b$ in the  interval $0 \leq \b  \leq \q $
    (with
$a=\sqrt 3$  and $a=1$ Melvin  thus  being the boundary points).

For generic values of $\q  $ and $\b  $ all of the type II
supersymmetries are broken  \refs{\gib,\magnetic}.
 Supersymmetries are restored at the
 special values $\q R=2n_1$ and $\b  \td R=2n_2$ ($n_i=0,\pm 1, ...$),
where the conformal model describes the
standard flat-space  type II  superstring theory (see below).

According to \fff, a shift of $x_9$ by period of the circle
$2 \pi R$  implies rotation in the plane by angle $ 2 \pi b R $.
Thus in  the special case of $b R=2k +1 $ ($k=0, \pm 1, ...$)
the metric becomes topologically trivial. However,
the superstring theory
on this background is still non-trivial (not equivalent to
that of standard flat space)  since space-time fermions
change sign under $2\pi$ rotation in the plane. That means that the $b R=1$
case (all models with $b R=2k +1$ are equivalent)
  describes superstrings with {\it anti}periodic
boundary condition in $x_9$, just as in the
 twisted 3-torus model        of \rohm\
or in the
finite temperature case \atick\ (see also \koun).
As was  already discussed  in section 1,
the type IIA string  in the space \fff\  with $\q R=1$ is
equivalent, for $R\to 0$,
to type 0A string
 in  $R^{1,8} \times S^1_{R\to 0}$ or type 0B string in
$R^{1,8} \times S^1_{\tilde R\to \infty}$.
 The tachyon of type 0 theory
  originates from  a particular   winding mode
 present in the $\q R=1$  type II model
 spectrum (see section 2.1).

\subsec{\bf String spectrum}

Despite the fact that the  $\sigma $-model \lagg\
(supplemented  with  fermionic terms given in  \magnetic)
 looks non-linear
and represents a curved  10-d background \melv,\
the corresponding superstring theory is exactly solvable
in terms of free fields \magnetic.
 Using  operator quantization
the   resulting string  Hamiltonian
 and the level matching condition are  given by \magnetic
   $$
\hat H =
 \ha  \a' ( -E^2 +  p_s^2  ) + \hat N_R+  \hat N_L +\   \ha{ \td R R\inv }
(m- \q R\hat J)^2
$$ \eqn\hail{
+ \ \ha { R  \td R\inv }  (w  -  \b \td R  \hat J)^2
-   
 \g  (\hat J_R-\hat J_L)=0 \ ,
}
\eqn\coss{
\N_R-  \N_L = mw  \ ,
}
where
\eqn\gam{
\ \ \ \ \g\equiv  \q Rw + \b \td R m
-  \q  R \b \tilde R \hat J \ .
}
Here $p_s$  are continuous  momenta in the 6 free directions,
and the integers $m$ and $w$ represent  quantized
 momentum and winding numbers in the compact $x_9$ direction.


$\hat N_R$ and $\hat N_L$ are the   number of states operators, which
have the standard free string theory
form in terms of normal-ordered bilinears of
bosonic and fermionic oscillators,
\eqn\coo{ \hat N_{R,L}= N_{R,L} -a \ ,
\ \  \
\ \ \    a^{\rm (R)} =0\ , \ \ \    a^{\rm (NS) } =\ha \ .
}
$\hat J\equiv \hat J_L + \hat J_R $ and
$\hat J_{L,R}$ are the angular momentum operators in the 2-plane
(i.e.
 constructed out of
oscillators corresponding to the 2-plane coordinates $X=x_1 + ix_2=
 r e^{i \vp}$)
with the
 eigenvalues
\eqn\eig{
\hat J_{L,R}
 = \pm (l_{L,R} + \ha) + S_{L,R} \   ,  \ \ \ \ \ \ \ \ \ \ \ \ \ \
 \hat J= l_L-l_R + S_L + S_R\ .
}
The orbital momenta $l_{L,R}=0,1,2,...  $ (which replace the continuous
linear momenta $p_1,p_2$ in the $(r,\vp)$ \ 2-plane
for non-zero values of $\g$)
 are  the analogues of the Landau quantum number,
 and $S_{R,L}$ are the  spin components
  (in the NS-NS sector, $|S_{R,L}|\leq \hat N_{R,L}+1$).
  In the special case of $\gamma=0$ (or, more
generally,  $\g=n$)
  the zero-mode structure changes in that the translational invariance
in the 2-plane is restored. This  leads to  a
modification in the above formulas, with
the orbital momentum parts becoming   the standard
flat-space theory expressions
  in terms of the  zero-mode operators
$x_{1,2}, p_{1,2}$.

The expression for the Hamiltonian
  \hail\ applies for $-1< \gamma < 1$,
while for other  values one is to replace $\gamma $ by
 $\hat \g\equiv \g - [\g]$, where $[...]$
 denotes  the integer part.\foot{This follows
 from a change in the normal ordering constant, see \exactly.}
The mass spectrum is  then   invariant under
\eqn\inve{  \q\to \q+ 2n_1 R\inv\ , \  \ \ \ \
\ \ \ \     \b\to \b + 2n_2\tilde  R\inv\ , \ \ \ \ \   \ \  n_1,n_2 =0, \pm
1, ... \ . }
This  becomes clear  by observing that
this transformation  may  be compensated
 by integer shifts of  quantum numbers.
 Note that  because  $\hat J$ can take
both
integer
(in NS-NS, R-R sectors) {  and}
{\it half}-integer  (in NS-R, R-NS sectors) values,
the symmetry of the
bosonic
part of the spectrum  $\q\to \q+ k_1 R\inv, \  \b\to \b+ k_2
 \tilde R\inv $  is {\it not}
 a   symmetry of  its
fermionic part, i.e. the full  superstring spectrum is
invariant under
\inve\  with  even integer shifts  $k_i=2n_i$ only.

In the case  of $\q R =2n_1, \ \b \tilde R = 2 n_2 $ (i.e.
$\g=$even integer, $\hat \gamma=0$) the
 spectrum is thus equivalent to that of  the standard  free superstring
compactified on a circle.
In the two cases
\eqn\dve{
(a)\ \
 \q R={\rm odd}, \  \ \b  \td R= {\rm even} \ \  \ \ \ \
 {\rm or }\  \ \ \ \
(b)\ \
  \q R={\rm  even},\  \  \b  \td R={\rm  odd } \ ,
}
the spectrum
 is
 the same as that
of  the  free superstring compactified
on a circle with antiperiodic boundary conditions
for space-time fermions \refs{\rohm,\atick}.\foot{This relation  becomes
particularly  clear
 in the light-cone GS formulation given in \magnetic. For
 example,  for $\b=0$ the  GS
 action  is quadratic in fermionic 2-d fields which are coupled
to  a  flat but topologically non-trivial connection.
 After a formal redefinition, one gets free action  in terms of
 fermion fields obeying  the  boundary condition
 $
 \Sigma _{R,L }(\tau,\sigma  + \pi ) =
 e^{ i \pi \g   } \Sigma_{R,L}
 (\tau,\sigma )$.  This makes it clear that the
 inequivalent  values of $\gamma$ are $ -1< \gamma \leq  1$,
 with $\gamma=1$ corresponding to the  antiperiodic case.
More precisely, the
fermions are antiperiodic with respect to
$x_9$, but  as functions of the string world-sheet direction $\sigma$
their
periodicity is determined by $\gamma= b R w$, i.e. they
are
periodic in $\sigma$ in  the even winding
sector  and antiperiodic in $\sigma$
in the odd winding sector.
}

 The T-duality symmetry in the compact Kaluza-Klein direction $x_9$
 (which exchanges the axial-vector  and vector magnetic
  field parameters  $\b  $ and $\q  $, cf. \baag)
is manifest in \hail: the Hamiltonian is
invariant under  $R\leftrightarrow \td R\equiv \a'R\inv ,\
 \q  \leftrightarrow \b,
\ \ m\leftrightarrow w$.

A close examination of \hail\ shows that   all states with
$\hat J_R-\hat J_L\leq \hat N_R +\hat N_L $ have positive (mass)$^2$.
The only bosonic states which may  be {\it tachyonic }
  thus lie on
the first Regge trajectory with maximal value of $S_R$, minimal value of
$S_L$,  and   zero orbital momentum,
i.e.  $\hat J_R=S_R-\ha=\hat N_R+ \ha ,\   $  $\hat J_L=S_L+\ha=-\hat N_L -\ha
$, so that $\hat J_R-\hat J_L= \hat N_R +\hat N_L +1$.
In particular, all fermionic  states
  have  (mass)$^2\geq 0$, as expected in a unitary theory.

{}From eq. \hail\ one learns  that in general there are
instabilities (associated with states with  high spin and charge) for
arbitrarily small values
of the magnetic field parameters.
The only exception is the   $\b =0$  model  (or its T-dual
 $\q =0$ model):  here the type II superstring
  model  has no tachyons if  the value of $\q$ is smaller than
  some {\it finite} critical value $\q _{\rm cr}$.
The spectrum of the $\b =0$ model is given by  ($M^2 = E^2 - p_s^2$)
\eqn\abajo{
\a' M^2= 2(\hat N_R +\hat N_L) +\a' (m R\inv - \q \hat J)^2 +
 {w^2 R ^2\over
{\a'}}
- 2 \hat \gamma  (\hat J_R-\hat J_L)\ ,}
\eqn\abbb{
   \hat \gamma =\gamma - [\gamma] \ , \ \ \ \ \ \ \ \ \ \ \ \
\gamma = \q  R  w \ . }
Here
the first (lowest mass)
potentially tachyonic state  that appears
  as $\q $ is increased from zero   has
\eqn\bajo{
\hat N_R=\hat N_L=0\ ,\ \ \ \
\hat J=0 \ ,\ \ \ \ \hat J_R-\hat J_L=1\ ,\ \ \ m=0\ ,\ \ w=1\ ,
}
$$
l_R=l_L=0\ ,\ \ \ \ S_R=1\ ,\ \ \ S_L=-1\ ,
$$
so that (for $\q R <1$)
\eqn\masa{
\a' M^2= {R^2\over\a' } - 2\q R \ \equiv  \ 2R ( \q _{\rm cr}  -\q)   \ ,
\ \ \ \ \ \ \   \ \q _{\rm cr}= {R\over 2\a'} \ .  }
It    becomes tachyonic for $\q >\q _{\rm cr}$.
Because of the periodicity   of the spectrum \inve\
 (cf. \gam,\hail\ for $\b=0$)
 the parameter $\q R$
may  be restricted to the interval
\eqn\res{
 -1< \q R \leq  1 \ . }
 Then  the stability
condition  $\q < \q _{\rm cr}$
implies  that  the theory with
\eqn\cri{ R\geq R_{\rm cr}\equiv   \sqrt{2 \a'} }
is tachyon free.

In  \masa\ we assumed that $\q R < 1$.
For $\q R=1$, one gets  $\hat \gamma=0$ so that  the  last
gyromagnetic  term in \abajo\ and thus the $- 2 \q R$ term
 in \masa\ is  absent.
However, the spectrum is continuous at this point:
there is an   extra negative term  coming from a
change in values of normal ordering constants
in $\hat N_R$ and $\hat N_L$ \magnetic, which  produce
  extra $-2$ term in $\a' M^2$ in  \masa\
  (see also below).

Similar  considerations apply to the  T-dual model with $\q =0, \
\b\not=0  $, which has the equivalent
 spectrum.
For this model there are
tachyons  for  any $\b  >\b _{\rm cr }$,
with $\b  _{\rm cr }= {{ \td R} \ov 2 \a'} = {1\over 2R }$.
If we  consider the region $ 0 \leq \td \q \td R < 1 $ then
this model is tachyon-free if $\td R\geq  \sqrt{2 \a'}$, i.e.
$R\leq   \sqrt{\a' \over 2}$.

In order to  interpret the spectrum  which follows from
 \hail\  in terms of  gyromagnetic interaction, i.e. to make explicit how
the  momenta  $\hat J_R$ and $\hat J_L$ couple to the magnetic field,
it is convenient to introduce
the ``left" and ``right" magnetic field parameters
$B_{L,R}\equiv \q \pm \b $  \ and charges
 $Q_{L,R}=mR^{-1}\pm w \td R\inv  $. The mass formula \hail\
 then
 takes the form
\eqn\mmfor{
 M^2 = M_{\rm free} ^2 -2 (B_LQ_L\hat J_R+B_R Q_R\hat J_L)+(
 B_L^2 \hat J_R + B_R^2 \hat J_L \big) \hat J\  ,
}
\eqn\mem{
 M_{\rm free}^2\equiv {2\over \a '} \big( \hat N_R +\hat
 N_L \big) +\ha  (Q_L^2+Q_R^2)\ .  }
In the model with $\b  =0$ one has only the  vector field background
(see \baag), i.e.
$B_L=B_R$. Then,  as is clear from \mmfor,
 the  gyromagnetic coupling of the  winding states
involves
the combination $\hat J_R-\hat J_L$, while  the coupling of the
momentum states -- the combination  $\hat J_R+\hat J_L$.
In the T-dual  model with $\q  =0$ one has  the axial-vector
field background, $B_L=-B_R$, so that the momentum states
couple to $\hat J_R-\hat J_L$ and the winding states to $\hat J_R+\hat
J_L$. In both cases, the states which first become tachyonic
have  gyromagnetic coupling  involving
 $\hat J_R-\hat J_L$.

The partition function of the model   found  in \magnetic\
exhibit the same symmetries as the spectrum:
\eqn\simm{
Z(R,\q ,\b )= Z(\td R, \b , \q ) \ ,  \ \ \ \
}
\eqn\siml{
Z (R,\q ,\b )= Z(R, \q  + 2n_1 R^{-1},\b  +2n_2  \td R\inv )\ , \ \ \ \
n_{1,2}
=0, \pm 1, ... \ ,   }
where $\td R$ is the T-dual radius \pee.
Eq. \simm\ expresses the T-duality, while \siml\
is the   symmetry \inve\
 of the full conformal field theory.

In the special cases   when  both $\q R$ and  $\b  \td R $
are even, the partition function is identically zero, since  for these
values  the theory is equivalent
to the free superstring theory with $\q =\b =0$.
Two other simple models which are
equivalent
  to the
free superstring theory with fermions obeying
{\it antiperiodic} boundary conditions in the $x_9$ direction
have parameters  given in \dve.
For non-integer  $  \q R  $ and $\b \tilde R$
there are tachyons for
 any value of the radius $R$, and so
 the  partition function contains infrared divergences.
   When  $ \b \td R$   is
 equal to  an even
 integer, $Z$ in \siml\
reduces to that of the $\b  =0$ (K-K Melvin)
  model, and is finite in the range of
  parameters  for which  there are no tachyons
  in the spectrum.
  Similar statement applies in the T-dual
  case of $\q R= 2 n$.
When either $ \b \tilde R= \a' \b R\inv $  or
$\q R$ is an odd number, the partition
function is also
finite in a certain range of values of the radius.

As was already mentioned above, the type IIA(B)
 theory in the background \fff\ or  \melv,\meel\
 with $\q R=1, \b=0$  is
 equivalent
 to the orbifold of type IIA(B)
 on  $(S^1)_{2R}/[(-1)^{F_{s} }\times {\cal S}]$,
 i.e. to the theory interpolating between type IIA(B)
 in  $R^{1,8} \times (S^1)_{R\to \infty }$
  and type 0A(B) in $R^{1,8} \times (S^1)_{R\to 0}$.
  In this case
 $\hat \gamma =0$, and the spectrum \abajo\ coincides
 with the spectrum of free superstring theory
with antiperiodic fermions.
As well known, in this case
tachyons may  appear in the  $w=$odd sector, because of the
reversal of
the GSO condition \atick .
The scalar state with  winding number $w=1$ and $N_R=N_L=0$ (i.e. with
 $\hat N_R=\hat N_L=-\ha$)\foot{This is transparent in  the GS formulation:
in the  odd winding (antiperiodic)
sector the operators $N_{L,R}$ have
half-integer eigenvalues
and the  normal-ordering constants are  equal to -1/2.}
is not projected out by the GSO, and has  the mass\foot{The extra (twisted
sector)
 set of R-R massless states of type 0 theory
also originate from the odd winding sector  with
$\hat N_R=\hat N_L=0$.}
\eqn\rios{
M^2={R^2\over {\a' }^2}- {2\over\a ' }\ .
}
In the limit of $R\to 0$,
it  represents  the only tachyonic state that
  is present in the spectrum of
 type 0B(A)   theory in ${R}^{1,9 }$.
It is interesting to see how this  negative mass shift
 emerges as the magnetic field is increased from zero.
 The tachyon corresponds to the state
\bajo, whose mass \masa , for
 $\gamma= \q R=1-\epsilon = \hat \g$,
  reduces to \rios\ in the limit $\epsilon\to 0$.
  In that sense  type 0 tachyon   corresponds to the
  special ``spinning" winding mode
  of type II theory in the background \fff,
  and its  mass $ - {2\over\a ' }$
  has its origin in the gyromagnetic interaction term.

  \bigskip

Let us note in passing  that some
of the expressions
 for the mode expansions  and the spectrum
 in this magnetic model
(see  \refs{\constant,\exactly,\magnetic })
 formally are  very  similar to the corresponding ones  for the
open strings in the presence of a constant
magnetic ($F_{mn}$ or  $B_{mn}$)
 field   \refs{\callan}. One may  think,
 by analogy with the open string case \sw,
 that one can then use certain limits of
 these magnetic flux tube models  to relate {\it closed} strings
 to  non-commutative field theories.
However, this does not seem to be possible.
The point is that  open strings
in  constant NS-NS $B_{mn}$ field   behave as  electric dipoles
with total charge equal to zero, so that
there is a residual translational  invariance and
non-zero commutators for
the center of mass coordinates.
In contrast,
closed string states can have only charges associated with
momentum or winding number, i.e.
  they do not behave like  electric dipoles.
For charged closed strings with $\g \neq n$, there is no
translational
invariance
in the $x_1,x_2$ plane (there are Landau levels instead)
and  there is no obvious analog of the
non-zero commutator $[x_1,x_2]$  which appears
in the open string case
(one may  introduce operators $x_1,x_2$ formally defined in terms of Landau
level creation/annihilation operators,
which in the zero field limit reduce to the standard
 center of mass coordinates,
but they commute).

\subsec{\bf
 Tachyon  spectrum  from  effective field theory}

The  full string mass spectrum  \hail\ contains, in particular,
the spectrum of small fluctuations of
 type II supergravity  fields expanded near the  curved background
 \melv,\meel.  To decouple string modes one is to  take
 $\a'\to 0$  (assuming that $\sqrt {\a'}/ R, \sqrt {\a'}\q ,
  \sqrt {\a'} \b \to 0$).\foot{Since  there is  a
  compact $x_9$ direction, one may try to
go further  and
consider, in addition to $\a'\to 0$,
 the  low-energy approximation  that ignores   not only
the  massive string  modes  but also  the Kaluza-Klein modes.
Let us note that  the  Kaluza-Klein truncation is {\it not}
 a valid approximation near $\q R=k_1$, or
$\b \td R =k_2$ (for integer $k_1,k_2$), where  the theory is essentially
ten-dimensional.
 This can be seen from the above mass spectrum \hail,\mmfor.
For example, consider
 the case  of $\b  =0$ and the non-winding supergravity
 sector ($\hat N_R=\hat N_L =0$,
 \ $w=0$),
so that $ \gamma =0$ (see \gam). Then the  mass formula \abajo\  becomes
 $M= R^{-1}(m-\q R\hat J)\ , $
so  that when $\q R$ is near 1,
there are Kaluza-Klein states which become light. These are the
states with
$m=\hat J$ which have  mass
 $M= {m \over R_{\rm eff} }\ , $
 \
$ R_{\rm eff} \equiv {R\over 1-\q R}\ .$
The Kaluza-Klein approximation thus
fails when $R_{\rm eff} $ is large.}

  In the model with $\q =0$ \beb\  the first state
that becomes tachyonic as the field $\b  $ is
increased from zero is from  the ``massless"
sector of the theory, i.e.  it has  $\hat N_R= \hat
N_L=0$ and $w=0$,  $S_R=1,\ S_L=-1$ (cf. eq.~\bajo\ in T-dual theory).
This implies that one should be able to
detect  this instability
directly from
the effective supergravity field equations, i.e. without
  solving first
the conformal string  model.
  It is useful to see how this happens explicitly,
   since similar supergravity-based
 considerations may apply in other related cases
  (magnetic R-R backgrounds, or magnetic
 backgrounds in $d=11$) where  one does not
  have the  benefit of knowing the exact
 microscopic (string theory   or M-theory)  spectrum.

Consider   the general  magnetic flux tube
model \melv,\meel\
 with both $\q ,\b \neq 0$.
In the sector $\hat N_R= \hat N_L=0$, $w=0$  the mass formula \hail\
 becomes
\eqn\secc{
M^2=( { m\over R}-\q  \hat J)^2+  {\b ^2 } \hat J^2
+ 2 \b ({m\over R}-\q \hat J) (l_R+l_L+1-S_R+S_L)\ ,
}
$$
\hat J=l_L-l_R+S_L+S_R\ ,\ \ \ \ \ \ \ -1\leq S_{R,L} \leq 1\ .
$$
Assuming  $ \b ({m\over R}-\q \hat J)= \gamma/\a' >0$
(see \gam),
 $M^2$ can become  negative,  provided $S_R=1$, $S_L=-1$,
and $l_L=l_R=0$.

Let us now  try to reproduce  this spectrum
of  instabilities from   the effective field theory.
It is useful to start with   a scalar
supergravity  fluctuation mode $\psi  $
that satisfies the ``massless" wave  equation in the
 background \melv,\meel,
\eqn\msa{
\p_m(  e^{-2\phi}\sqrt{G} G^{mn}\p_n)\psi  =0
\ . }
Such spinless  mode is not  expected to become tachyonic
since,  according to \secc,   the tachyonic state should have
a non-zero  spin,
but we shall see that \msa\ does reproduce
 the expected  spectrum  \secc\ in the sector $S_R=S_L=0$.

Using  \melv,\meel ,  eq. \msa\ takes the form
$$
\big[-\p_t^2+\p_{x_s}^2+{1\over r}\p_r (  r\p_r)  + {1\over r^2}(1+ \q ^2r^2)
(1+\b ^2 r^2)\p_\varphi^2
$$
\eqn\mass{
+\ (1+\b ^2 r^2)\p_{x_9}^2 -2 \q (1+\b ^2 r^2)\p_\varphi \p_{x_9} \big]
\psi =0\ .
}
Parametrizing  the eigen-mode  as
\eqn\phh{
\psi = e^{i E t +ip_s x_s + ip_9 x_9 + il \varphi }\eta (r)\ ,\ \ \ \ \ \
\ \ p_9={m\over R}\ ,\ \ \ \ \ m,l= 0, \pm 1, ...\ ,
}
we get from ~\mass\
\eqn\ttt{
\big[- {1\over r}\p_r( r\p_r) +{l^2\over r^2} +\nu^2 r^2 \big]
\eta (r) =\mu^2 \eta (r)\ ,
}
where
\eqn\eee{
\nu = \b   (p_9-\q  l)\ ,\ \ \ \ \
\mu^2= M^2 - (p_9 -\q l )^2 -\b  ^2 l^2 \ ,\  \ \ \ \ \ \ \
M^2\equiv E^2-p_s^2\ .
}
This  is the  Schr\" odinger equation for a two-dimensional
 oscillator with the frequency $\nu$.
It has normalizable solutions only if $\mu^2= 2\nu (l_L+l_R+1)$, i.e.
\eqn\spec{
M^2_{\rm spin \ 0}  =  (p_9 -\q l )^2 + \b  ^2 l^2  +
  2\b  (p_9-\q  l) (l_L+l_R+1)\ ,\ \ \ \ \ \ \
   \ l_L, l_R=0,1,2,...\ .
}
Here $l_L,l_R$ are related to the Landau level $l$ and the
radial quantum number $k$
 by
\eqn\rell{
l_L-l_R=l \ , \ \ \ \ \ \ \ \  l_L+l_R=2k+|l| \ . }
The mass spectrum  \spec\  coincides indeed
with  \secc\  after one sets there  $S_R=S_L=0$,
as appropriate for a scalar fluctuation.
This scalar mode has $M^2\geq 0$.

The mode that  becomes tachyonic  should have, according to \secc,
the non-vanishing spin  in the 2-plane,
 $S_R=1$, $S_L=-1$.
Deriving the  analog of eqs. \msa,\mass\
 for such  fluctuation   directly  from the supergravity  equations
 is somewhat  complicated.\foot{The unstable mode should
 be a particular mixture of the
  metric  and the antisymmetric 2-form
   perturbations. There is some analogy with   non-abelian  gauge theory
 where  the tachyon mode in a constant  magnetic background
 has spin 1 \niel.}
 We shall by-pass this  step  by making a natural conjecture
 that to generalize \spec\ to the case of
  non-zero spin of the fluctuation
 one is simply  to  add the spins $S_{L,R}$
  to the orbital momenta $l_{L,R}$
 in \spec.
 This conjecture is indeed confirmed by the expression
 \secc\ following from string theory.

 In the case of interest ($S_R=-S_L=1$)  we obtain
\eqn\speci{
M^2
 = (p_9 -\q l )^2 +\b  ^2 l^2 +
 2\b  (p_9-\q  l) (l_L+l_R-1)\ ,\ \ \ \ \ \ \ \   l_L, l_R=0,1,2,...\ .
}
As was already
mentioned above, and as is  now   clear   from \speci ,
 the first tachyon that  appears  for $\b  >\b  _{\rm cr}$
 ($\b _{\rm cr}={1 \ov 2R}$)
 is  the  state with $l_R=l_L=0$.

Let us   note that in addition to
the perturbative instability discussed above, the
  magnetic backgrounds in
{\it  supergravity}  suffer also
from a non-perturbative instability  -- a
decay into  monopole-antimonopole pairs
\refs{\gib,\gaun}. This instability is  present for any value
of the magnetic field, i.e. even when perturbative
 instability is absent.

\newsec{$d=10$  supergravity backgrounds with  Ramond-Ramond magnetic fluxes
and their $d=11$ counterparts}

By  using U-dualities, it is easy to convert
the  NS-NS magnetic
background \melv,\meel\  into  $d=10$ type II supergravity
backgrounds
with  magnetic fluxes  from the R-R  sector.
For example, starting with
\melv,\meel\ with $\b\not=0$
(i.e.  with non-vanishing 2-form field)  considered
as solution of  type IIB supergravity
and applying S-duality we obtain another
solution   with non-vanishing
R-R  2-form field (i.e. with the R-R axial-vector
field in $d=9$, cf. \baag).

In contrast to the NS-NS model \lagg, it is  not clear how to solve
the corresponding R-R superstring models:
the Green-Schwarz action in the light-cone gauge \green\
remains non-linear and is not related
to a free theory by  T-dualities and coordinate shifts,
 as was the case for \lagg\   \magnetic.

\subsec{\bf  Magnetic flux  7-brane of type IIA theory}

Another   example of R-R magnetic  background
 is provided  by  the type IIA solution
related  to the  $\b=0$  NS-NS  Melvin  background \melv,\meel\
of type IIA theory by the
chain of duality transformations: T$_{x_9}$ST$_{x_9}$.
This gives  the following background which may be interpreted as
a
non-supersymmetric R-R magnetic flux
 7-brane  of  type IIA theory \green\
\eqn\mccc{
ds^2_{10}=f^{1/2}\big( -dt^2+dx_s^2 + dx_9^2 +
dr^2 + r^2 f\inv d\vp ^2 \big)\ ,
}
\eqn\mzzz{
\ \ \  e^{{2}\phi }= e^{2 \phi_0} f^{3/2} \ ,\ \ \ \ \ \ \ \ \
A_\vp =\q r^2   f\inv \  \ , \ \ \ \ \ \ \
\ \ \ \ \ f \equiv 1+ \q ^2 r^2 \ ,
}
where $A$ is the R-R 1-form with the field strength
\eqn\fffw{
F=    { 2 \q    \ov (1 + \q ^2 r^2 )^2 }   r dr \wedge d \varphi  =
 2 \q  f^{-2}  e^r \wedge e^\vp \ ,}
which is approximately constant near the core.
The 7-brane directions are $x_s (s=1,..,6)$ and $x_9$.
Like the $d=9$ reduction of
 K-K Melvin  background \gaun, this
  solution can be obtained also directly  \refs{\green}
by dimensional reduction  of  the
$d=11$ background
with trivial 3-form field $C_{\mu\nu\rho}=0$
and the   metric (cf. \fff)
\gaun\
\eqn\maaa{
ds^2_{11}=-dt^2+dx_s^2 + dx_9^2
+dr^2+ r^2 (d\vp +  \q\  d\x )^2+dx^2_{11} \ .
}
Here $\varphi \equiv \vp +2\pi $ and $\x $
has period $2\pi  R_{11}$,  so  this  metric is
topologically non-trivial if
$\q R_{11}\not=n$.
Since
the metric is locally flat this  is an exact solution to the
equations of motion of $d=11$  supergravity {\it and} M-theory
(all possible higher-order curvature corrections vanish).

If the $11\to 10$  dimensional reduction is
done along the  coordinate $x_9$
  one obtains  the $d=10$
  NS-NS  background  \melv,\meel\ with $\b  =0$ or \fff\
   (K-K Melvin),
  discussed in
\refs{\gaun,\nucleation }.
 Dimensional reduction   along
$\x $  gives  the  above type IIA solution \mccc,\mzzz\
with  the magnetic R-R  vector  field.
The direct $d=11$ lift of the NS-NS  background \fff\
 is related to \maaa\ by the 9-11 flip:
\eqn\meee{
ds^2_{11}=-dt^2+dx_s^2 + dx_9^2
+dr^2+ r^2 (d\vp +  \q \ dx_9 )^2+ dx^2_{11} \ .
}
 If  the effective string coupling $e^{ \phi}$ in \mzzz\  is chosen to be
  small  at  $\q r\ll 1$ ($ e^{\phi_0} \ll 1$)  it  gets
  large at $\q r\gg 1$, so  one may say
  that  the geometry is ten-dimensional
  close to the core of the
   7-brane (inside the flux tube)  but  becomes
effectively eleven-dimensional  far  from it \refs{\green,\CG}.

The remarkable circumstance  that the R-R 7-brane \mccc,\mzzz\
 is a dimensional reduction of
 locally-flat $d=11$ background  \maaa\ was utilized
 in \green\ to derive the exact
expression for the  corresponding  $d=10$
GS superstring action
 starting with  the known  flat-space
 $d=11$ supermembrane action \bst.

It is interesting to note a   similarity between the
flux 7-brane \mccc,\mzzz\ and D6-brane   of
 type IIA theory -- both backgrounds  have only the dilaton and
 the magnetic R-R vector as non-trivial matter
  fields.\foot{In a sense,
   this flux-brane may be
 viewed as ``connecting" two  magnetic R-R charges  --
 D6-brane  and anti D6-brane (cf.
 \refs{\nucleation,\CG}).}
 The dependence on the function
   $f(r)$ in the metric \mccc\ and the dilaton field
 \mzzz\ is the same as the dependence on  the inverse
 of
  the harmonic function $H^{-1}= (1 + { L\ov \rho})^{-1}$
 in   the D6-brane solution.
 This formal correspondence
 implies  that applying T-duality along  the  ``parallel"
 directions ($x_1,...,x_6,x_9$) of the 7-brane  one
 obtains similar  flux p-branes  having the same type of R-R
 fields  and the same structure of $f$-dependence
 as in the  Dp-brane  solutions of IIA/B theories.
 In particular, we  get the following type IIB
 flux 4-brane  solution, which,
 like the  D3-brane background, has {\it constant}  dilaton. It has
  the   metric
 \eqn\mccc{
ds^2_{10}=f^{1/2}\big( -dt^2+dx_1^2 + ... +dx_4^2  + dr^2)
 + f^{-1/2} (  dx_5^2 + dx_6^2  +  dx_9^2
  + r^2 d\vp ^2 \big)\ ,
}
and  the  self-dual  R-R  5-form  with   magnetic
 component $F_{r\vp 569}\sim b f^{-2}$.

\subsec{\bf   ``Mixed" type IIA 2-parameter  magnetic background  }
\def \y {{x_9}}

Starting with the general  NS-NS  solution
\melv,\meel\   and  applying again T$_{x_9}$ST$_{x_9}$ dualities
 one obtains   a
generalization of the R-R  Melvin solution \mccc , \mzzz\
in type IIA theory.
It  depends on two
  parameters which  control the strengths of the R-R magnetic
1-form   field $A$ and the  NS-NS  2-form field $B_2$:
\eqn\dosa{
ds^2_{10A}= f^{1/2}\big( -dt^2+dx_s^2+ {\tilde f\inv  }dx_9^2 + dr^2 +
{ f\inv  \tilde f\inv } r^2  d\varphi^2\big)\ ,
}
\eqn\aave{
A  = \q  r^2 f^{-1} d \vp \ ,\ \ \ \ \ \ \ \ \
 B_2 =\b  r^2 \tilde f^{-1}  {d \varphi\wedge d x_9 } \ ,\ \ \ \ \ \
\ \ e^{2\phi }=e^{2\phi_0 } f^{3/2}\tilde f\inv  \ ,
}
$$
f=1+\q ^2  r^2\ ,\ \ \ \ \ \ \ \ \ \ \tilde f=1+\b  ^2 r^2\ .
$$
 The R-R 7-brane background \mccc ,\mzzz\ is
the special  case of $\b  =0$.

Lifting this type IIA
background to $d=11$ we  find the following
``6-brane" solution of $d=11$ supergravity
\eqn\surr{
ds^2_{11}= \tilde f^{1/3} ( -dt^2+dx_s^2)
 + \tilde f^{-2/3} \big[ dx_9^2 +  \tilde f  dr^2  + r^2  f\inv  d\varphi^2
+  {f }\big( d\x +{\q r^2 f\inv } d\varphi\big)^2 \big]
\ , }
\eqn\ccf{
C_3 = \b  r^2 \tilde f^{-1} d x_{11} \wedge d x_9 \wedge d \vp   \ .
}
It   reduces to the previous
flat  background \maaa\ when $\b  =0$.
This $d=11$ background is related by the 9-11 flip to
the background obtained by  lifting the NS-NS  solution
 directly to $d=11$ along $x_{11}$.

Another special case is $\q=0$, when the metric \surr\ becomes
\eqn\surre{
ds^2_{11}= \tilde f^{1/3} \big( -dt^2+dx_s^2  \big)
 + \tilde f^{-2/3} \big( \tilde f dr^2  + r^2    d\varphi^2
 + dx^2_9  +   dx^2_{11} \big)
\ . }
Remarkably,  this metric (and $|C_3|$ in \ccf)
are  {\it symmetric}
 under the ``9-11" flip.\foot{This metric  is
regular:
its  Ricci scalar is $ R={2\b^2\ov 3(1+\b^2 r^2)^{7/3}} , $
\  $R^2_{mn} = 23 R^2$,
and  $R^2_{mnkl}$ has similar behavior.}
That means that compactifications along $x_{11}$ 
or along $x_9$ lead  to the same  NS-NS background:
 the $\q=0$ case of the
2-parameter  background \melv,\meel, i.e. to \beb.

By applying T- and S- dualities to \dosa,\aave\
it is easy  to construct other similar
solutions with mixed magnetic
fluxes.
Furthermore, it is possible to generalize these
$d=10$ and $d=11$ solutions to
the case of more than two   magnetic parameters.
Examples of such   more general solutions  will be
 presented in section  6.

\newsec{Tachyon instabilities of magnetic backgrounds
and type 0 tachyon}

As was mentioned in  section  2.1,
there are
two  magnetic  models \dve\
 which
are
equivalent
to  type II superstring  theory   in flat space
with fermions obeying {\it anti}periodic
boundary
conditions in the
$x_9$ direction \magnetic. Modulo periodicity  in
$\q R$ and $\b \tilde R$  \inve,\siml\
 they are  simply  ($R=R_9$)

(a) the model with $\b=0,\  \q R=1$ ; \ \ \ \ \ \ \ \
 (b) the model with
$\b
\tilde R=1, \  \q=0$.

\noindent
They are equivalent (T-dual) as  weakly-coupled  type II
superstring theories compactified on a circle.\foot{Note that while the
periodicity
  $\q R_9 \to  \q R_9  + 2n $  is obvious  in the background
  \fff , its counterpart   $\b\tilde  R_9 \to  \b \tilde R_9  + 2n $
  is not visible at the level of  the supergravity background \beb .   }
The direct  lift of the corresponding $d=10$
 solutions  \melv,\meel\  to
$d=11$ gives   M-theory  backgrounds
 \meee\  and \surre,\ccf.
 T-duality equivalence is believed to hold also for finite coupling, so one
expects  that M-theory  compactified  along
   $x_9,x_{11}$  in the
  background  \meee\  with $ \q R=1$ and in  \surre,\ccf\
  with  $\b
\tilde R=1$    should be equivalent not only in the weak-coupling
  limit $R_{11} \to 0$ but also for finite $R_{11}$.
Thus it is natural to conjecture that M-theory in the  background
 \surre,\ccf\ with $\b \tilde R_{9} =1$  compactified on
 the $(x_9,x_{11})$ torus  with $(+,+)$ (periodic)   boundary conditions
 for the fermions is equivalent to
 M-theory in flat space compactified
 on 2-torus  with the  $(-,+)$ fermionic  boundary
 conditions.\foot{It should be noted that
while the background \melv,\meel\ represents an exact
(tree-level) superstring
solution, its $d=11$ lift --
the curved $d=11$ background \surre,\ccf\
 may, in principle,
  be deformed (in contrast to the flat  background
\maaa)  by higher order M-theory corrections.}

  The  M-theory   description of $d=10$ type 0A theory  in  \CG\
was  based on  the $d=11$
 background \maaa\ with $\q R_{11} =1 $.
Now  we  are able   to  suggest
another description of type 0A string theory --
 as M-theory in  the
background
\surre,\ccf\  with  periodic $x_9$ and
$\b \tilde R_{9} =1$.
 The type 0 coupling is
 $g_{s0}={R_{9} \ov \sqrt{\a' }} $. 
 Because of the 9-11 symmetry
 of \surre,\ccf\
  one may  consider   also
the equivalent
model with $\b \td R_{11}=1$ and
$g_{s0}={R_{11} \ov \sqrt{\a' }}.$

While the reduction of  \maaa\ to 10 dimensions 
gives  string theory in R-R F7 background,  the reduction  of 
\surre,\ccf\  gives apparently much simpler 
string theory in  NS-NS background 
(\melv,\meel\ with $b=0$)
the spectrum of which (in  {\it weakly-coupled} regime)
   we know. 
However,  the 9-11 flip  necessary to  relate
 type II and type 0 theories implies that, e.g., a weakly coupled
($R_{11} \ll R_{9}$) type II theory is mapped into  a strongly
coupled  ($R_{9} \ll R_{11}$)  type 0 theory   or vice versa.
That  precludes one from drawing immediate conclusions 
about the  presence or absence of tachyons  in 
type 0 theory in the NS-NS Melvin background directly 
from the known weakly coupled string spectrum.
In particular, the weakly coupled type 0 theory in  the  
 NS-NS Melvin background is certainly unstable for any 
value of magnetic field parameter (because of its
own  tachyon  present already 
 in flat space)
but it is expected to become stable at strong coupling 
and critical magnetic field  since   it should then become 
 equivalent  to a weakly coupled type II string in 
NS-NS background with  a small magnetic field parameter.

According to  \CG , type IIA and type 0A theories in
 the $\b=0$  R-R Melvin
background \mccc,\mzzz\  are  equivalent,
being related by a  shift of the R-R field strength parameter:
$b_0= b - R\inv_{11} $.
The above discussion suggests the following generalization of this
relation  ($R=R_{11}$):
\eqn\ayay{\eqalign{
{\rm Type\ IIA\ in}\  (\q,\b )\ {\rm   Melvin}  &=
{\rm Type\ 0A\ in}\  (\q -R^{-1},\b )\  {\rm   Melvin}  \cr
&={\rm Type\ 0A\ in}\  (\q ,\b -\tilde R\inv ) \  {\rm  Melvin  } \
. \cr}
}
It was  conjectured  in \BG\
  that the type 0A tachyon originates from a
massive mode  (in twisted sector of  $\Sigma$-orbifold)
 in  microscopic M-theory,
i.e. a mode which is
absent in the $d=11$  supergravity spectrum.
Here we will
present  evidence
that in the  new M-theory
description of type
0A theory -- as M-theory in  the background \surre,\ccf\
at the critical magnetic field $\b \tilde R_9=1$ --
 the type 0A tachyon  corresponds  to
 a particular fluctuation mode of $d=11$  supergravity
 multiplet.
 Below  we shall investigate the appearance of
instabilities in these
 models,
 looking at  the  effective field
theory spectrum
in different regions of the parameter space.

\subsec{\bf Spectrum of $d=11$
fluctuations in magnetic backgrounds}

As we have seen  in  section 2.2, the leading
tachyonic instability  in the model
\lagg\ with  $\q =0$  is  associated with  a ``non-stringy" state
and so can be found directly in the effective low-energy
field theory, i.e. in the $d=10$ supergravity.
The same  tachyonic state  in the  T-dual  $\b=0$  model
is the winding state  \masa\  and thus it
can  not be seen  in the  supergravity approximation.
In the corresponding $d=11$ description
this string  state  should be represented by a winding membrane state.
In the   U-dual  R-R Melvin  model      \mccc,\mzzz\
  this tachyon  should  also be a winding string state
(as can be understood using the 9-11 flip).

Following the discussion of section 2.2, let us first
 study  some
examples
where the presence of  tachyonic states can be seen directly at the
level of
supergravity equations expanded  near the magnetic background.
Applying  U-duality (which is a symmetry of the supergravity
equations of
motion
\HT)
to an unstable (say, NS-NS)
supergravity background gives,  of
course,  a  supergravity solution which is also unstable.
This  allows one to identify a perturbative
tachyon state in the  R-R background
\dosa,\aave\ with $\q =0,\b\neq 0$.
Lifting a (potentially) unstable $d=10$  solution to $d=11$
 gives a (potentially) unstable $d=11$ supergravity
  background.

Consider the $d=11$ background  \surr ,\ccf\  with $\q =0$, i.e.
\surre.
Linearizing the $d=11$ supergravity  equations  near
this background one can
find the  corresponding Laplace-type equations for the bosonic
fluctuation modes.
In general, there should be
   scalar, ``spin 1",  ``spin 2"  fluctuations
   (we are referring to angular
    momentum in the $(r,\vp)$ plane).

Consider first the equation  for some massless  scalar
fluctuation $\psi$, i.e.   let us  assume,  as in \msa,
 that
 there  is  a fluctuation mode  that satisfies
\eqn\dzero{
\del_\mu (\sqrt{G} G^{\mu\nu}\del_\nu ) \psi =0\ .
}
For  the metric \surre\ this becomes
\eqn\ecua{
\big[ -\del _t^2 +\del _{x_s}^2 + (1+\b ^2 r^2)
(\del _{x_9}^2+\del _{x_{11}} ^2 +{1\over r^2}\del _\varphi^2
)+{1\over r}\del _r (r\del _r)\big]
\psi =0\ .
}
Setting  as in \phh\
\eqn\hhh{
\psi =e^{i E t+ip_s x_s + ip_9 \y +ip_{11}  \x + il \varphi }\eta
(r)\ ,\  }
$$
\ \ \
\ \ \ p_9={m\over R_9}\ ,\ \ \ \ \
\ p_{11} ={n\over R_{11}}\ ,\ \ \ \ \ \  m,n=0,\pm 1, ... \ ,  $$
we get
\eqn\ecu{
\big[ -E^2 +p_{s}^2 + (1+\b ^2 r^2) \big( p_9 ^2+p_{11} ^2
+{l^2\over
r^2}\big)
- {1\over r}\del _r (r\del _r)\big] \eta (r) =0\ .
}
This equation  is of the same form as eq.~\ttt, so its
 spectrum is
\eqn\robb{
M^2\equiv E^2-p_s^2 = p_9^2+p_{11} ^2 +\b ^2 l^2 + 2 \b
  \sqrt{p_9^2+p_{11} ^2}(l_L+l_R+1)\ ,\ }
  where
$  l=l_L-l_R=J $ (cf. \rell).
In the sector $p_{11}=0$
the spectrum  \robb\ coincides  with
the ten-dimensional counterpart
\spec\
with $\q=0 $.

To generalize this spectrum to the case of the spin 1 and spin 2
fluctuations
we  add (as in section 2.2)  the spin contributions $S_R$, $S_L$ to
the
orbital
angular momentum parts.
Indeed,
since the $d=11$  background \surre\
has   translational symmetry in $x_{11}$,
the fluctuation modes that do not depend on $x_{11}$, i.e. have
$p_{11}=0$,
must be the same as in  case of the corresponding $d=10$  solution
(eqs.~\melv,\meel\ with $\q =0$).
But the spectrum of the latter
 is known, being part of the  string theory  spectrum
 in the NS-NS Melvin
 background
\hail, with its
supergravity part
 given by eq.~\secc\ with $\q =0 $.
Thus the $d=11$ spectrum  should be
\eqn\sbbb{
M^2=p_9^2+p_{11} ^2 +\b ^2 J^2 + 2 \b  \sqrt{p_9^2+p_{11}
^2}(l_L+l_R+1-S_R+S_L)\ ,
}
$$
J=l_L-l_R+S_L+S_R\ .
$$
This  result should also   follow
directly from the detailed
analysis of the supergravity equations for the  graviton -- 3-form
 fluctuation modes.
  A  negative contribution to $M^2$  appears from the gyromagnetic
interaction
  for the supergravity
  modes  with   $l_L=l_R=0$, and $S_R-S_L=2$,
   i.e. with the  maximal  spin $S_R=1$ and the
   minimal  spin $S_L=-1$.
  For such states
\eqn\robbb{
\eqalign{
M^2 &=p_9^2+p_{11} ^2  - 2 \b  \sqrt{p_9^2+p_{11}^2} \cr
&=   m^2 R^{-2}_9  +    n^2 R^{-2}_{11}
  - 2 \b  \sqrt{ m^2 R^{-2}_9  +    n^2 R^{-2}_{11} }          \ .
\cr }
}
This mass spectrum  \robbb\
exhibits the presence of tachyons.
Without  loss of generality, we may  assume that $R_{11}< R_9$
(otherwise we would identify $\y$ as the eleventh dimension).
Increasing $\b$ from the  zero value (corresponding to the
stable flat vacuum),  the
first tachyon appears for $m=\pm 1$, $l=0 $ and $n=0$ at $\b  >\b
_{\rm
cr}$,
with
$\b _{\rm cr} = {1\over 2 R_9}$. This is,
of course,  the same tachyon that was present   in the $d=10$
model
\speci.
The first tachyonic state with non-zero $p_{11}$
  will be a state with $n=\pm 1$, $l=0=m$, at  the critical field
$\b _{\rm cr}'={1\over 2R_{11}}$.
Since $\b _{\rm cr} < \b _{\rm cr}'$,
this  instability is  physically irrelevant --
 there should  be a vacuum transition at smaller field, i.e.
before it could   appear.

Similar analysis  can be repeated for the supergravity part of the
spectrum  of  M-theory in the  background \meee.
However, to  establish  a
correspondence  with   \robbb\ implied by T-duality relation in
$d=10$,  here we are to look at the {\it winding}
 part of the spectrum,
i.e. need to go  beyond the supergravity approximation.
The result for  the relevant sector of the
spectrum of  the flat   $d=11$ model  \meee\
with  $\q\not=0 $  may  be obtained by  ``uplifting" to
eleven dimensions of
 the
spectrum \abajo\ of the T-dual NS-NS  string model
and re-interpreting it as  {\it membrane}  spectrum.
We will only consider the sector $\hat N_R=\hat N_L=0, \ m=0$,
where the spectrum \abajo\ is
\eqn\rccc{
M^2= {w^2 R_{9} ^2\over {\a'}^2} +
\q ^2 J^2 +  {2\q  w R_{9} \over\a'
}(l_L+l_R+1-S_R+S_L)\ .
}
Re-written in  terms of the  M-theory  parameters
\eqn\oopp{
\a '= (4\pi ^2  R_{11} T_2)^{-1}\ ,\ \ \ \ \ \ \
 T_2= (2\pi l_P^3 )\inv \ ,\   }
where $T_2$ is the membrane tension and $l_P
$ is the eleven
dimensional
Planck length,
eq. \rccc\ becomes
\eqn\rbbb{
M^2= (4\pi^2 wR_9R_{11}T_2 ) ^2 +\q ^2 J^2 + 8\pi^2 \q  w  R_9
R_{11}T_2(l_L+l_R+1-S_R+S_L)\ .
}
The first term represents the usual
contribution to $M^2$ coming from a
 membrane of tension $T_2$ wrapped on a
2-torus (wound $w$ times around $x_9$ and
once around $x_{11}$).
The spectrum is similar to \sbbb ,
as expected from the T-duality relation between the corresponding
ten-dimensional models.
The counterpart of the first two terms in \sbbb\ is now a winding
term;
the gyromagnetic interaction in \sbbb\ is traded
for  an analogous  gyromagnetic
term
where the charge is now
the winding number $w$.

In this way we determine
 part of the spectrum for the $d=11$ flat model \meee\
 which is relevant for
 the study of instabilities.
The  winding membrane with quantum numbers
$S_R=1=-S_L$, $l_L=l_R=0$  thus has the mass
\eqn\rddd{
M^2= (4\pi^2 w R_9R_{11}T_2 ) ^2 - 8 \pi^2\q  w  R_9 R_{11}T_2\ .
}
As
 the magnetic parameter $\q $ is increased from zero,
 there is a critical value at
 which $M^2$ for this   state  changes sign.

\subsec{\bf Instabilities in type 0 and type II
 theories}
\def\aa {\alpha'_{9}}

As discussed above, the  M-theory in the magnetic background
\surr,\ccf\ \ with  $\q R _{11} =1, \b=0 $ (model
(a))
 or  with
\eqn\anpp{
\q=0\ ,\ \ \ \
\b \td R_{11}={\b \aa \over R_{11}}= {\b l_P^3\over 2\pi R_9
R_{11}}= 1\ ,\ \ \
}
 (model (b)),
should describe type 0A string theory with parameters
\eqn\pram{
\a '={l_P^3\over 2\pi R_{11} }\ ,\ \ \ \ \ \ \ \  g_{s0}={R_{11}\over
\sqrt{ \a '} }\ .
}
In eq. \anpp , the parameter $ \aa  ={l_P^3\ov 2\pi R_{9}}$
represents the string scale for  type IIA string theory in the
magnetic
background
obtained by dimensional reduction in $x_9$ of the $d=11$ solution
\surr\
with $\q =0 $.

The weakly coupled type 0A theory contains a tachyon.
The problem of interest is to  identify
the corresponding state in the spectrum of M-theory in these
backgrounds,
and to study its evolution as a function  of the magnetic field
parameter.
It is also of interest to see if  other tachyonic
instabilities may appear in type 0A theory in a
different region of parameter
space. We will investigate these issues using the
expressions  for the spectrum given  above
in \sbbb , \rbbb .

The type 0A tachyon state has mass
$m^2_0 =-{2 \over \a'} ,\
\a'={l_P^3\over 2\pi R_{11} }.$
This mode is present  at least
in the perturbative type 0A spectrum, i.e.
for
$R_{11}\ll l_P $. The type 0B tachyon has the same
mass
and is present at least
in the region where the type 0B coupling is small, i.e.
$R_{11}\ll R_{9}$.

Now consider the description of type 0A string theory in terms of
the model (b)  based on \surre .
The symmetry of \robbb\
under $R_9\leftrightarrow R_{11}$
is related to the $SL(2,Z)$ isometry of the
background \surre,\ccf .
This  background may
 receive quantum corrections, but it is possible
 that
this  symmetry (being a $d=11$ isometry)  may   still be present  in the
quantum-corrected solution.
Then  the  M-theory in this background would
(as in the flat space \JS )
 have   the $SL(2,Z)$ symmetry.
In the picture of \BG ,
 the $SL(2,Z)$ symmetry of type 0B string theory
 may look  unexpected, since the directions
$x_9$ and $x_{11}$  are not equivalent, given
 different boundary
conditions for the  fermions.
 However, it was argued in \BG\  that the
perturbative type II
orbifold description of type 0B theory combined with the  9-11 flip
requires  $SL(2,Z)$ symmetry for consistency.

The present description based on model (b)
 may  be viewed as an additional
evidence for  the  S-duality of  type 0B theory, i.e. the
  symmetry under $ g_{s0B}  \to
g^{-1}_{s0B}$, with $g_{s0B}=R_{11}/R_9$.
Indeed, given the background \surre,\anpp ,
we are free to make dimensional reduction in $x_9$ or in
$x_{11}$. For $R_{11}\ll R_9$, one shows (as in section  2) that
 reduction in $x_{11}$ gives a theory
which is equivalent to the free superstring theory with
fermions  antiperiodic
 in $x_{9}$. This equivalence is
 expected to hold for any coupling,
in particular, for $R_{11}\gg R_9$.
But in this region, by reducing in $x_{9}$, one  shows  that
the theory is,  at the same time,
 equivalent to  the free superstring theory with
 fermions  antiperiodic in $x_{11}$.

Let us explore the $d=11$  spectrum \robbb\
 in the region $\b \td R
=1-\epsilon $, where
$\epsilon \to 0 $.
Dropping $O(\epsilon)$ term we get
\eqn\kkll{
M^2={m^2\over R_9^2}+ {n^2\over R_{11}^2} - {2\over\a' }
\sqrt{m^2+{n^2R_9^2\over R_{11}^2} } \ .
}
The state with  $m=0$ has mass
\eqn\kkkl{
M^2={n^2\over R_{11}^2} - {2\over\a' }
{nR_9\over R_{11}} \ .
}
This state is non-perturbative from the point of view of type 0
theory.
There is also a state with $n=0$, $m=1$, with
\eqn\aaww{
M^2={1\over R_{9}^2} - {2\over\a' }
\ ,
}
i.e. with the   same mass as  the type 0A tachyon in the
limit of large $R_9$.\foot{This mass formula is
 applicable for $R_{11}\ll R_{9}$, $R_{11}\ll l_P$,
or for $R_{9}\ll R_{11}$, $R_9\ll l_P$,
 given that  the condition \anpp\  is symmetric
under
the interchange of $R_9$ and
$R_{11}$.}
Thus in this description of type
0A string theory as M-theory in the background
\surre,\ccf\  it  is  consistent to identify the type 0A tachyon state
with a fluctuation mode of the $d=11$ supergravity multiplet.

Let us consider now  the
description of type 0A string theory as M-theory in the background
of \maaa\ of  model (a).
Setting  $\q R_{11} =1-\epsilon, \ \epsilon\to 0 $, the spectrum \rddd\ becomes
\eqn\reee{
M^2= {w^2R_{9} ^2\over {\a '} ^2}  - {2\over\a' }  {wR_9\over
R_{11} }
\  .
}
This state  may  be viewed as  ``T-dual"  of  the state
\kkkl .

In this model (a)
 description, the type 0A tachyon arises as the T-dual
counterpart of \aaww .
This can be understood
 by describing the type 0A theory as in section 2
by setting  $\q R_{9} =1-\epsilon  $.
We then get  the state with quantum numbers \bajo\ and
the  mass given by \rios .

For a general $\q $, this state has the mass given in \masa .
Using the duality relation \ayay , we can consider the correction
to the mass of the tachyon state in type 0A theory
when a magnetic field $\q_0 $ is turned on
\eqn\uuu{
M^2= {R_{9} ^2\over {\a '} ^2}  - {2\over\a' }(1-\q_0 R_9 )  \  .
}
As expected, this is negative  for small values of the magnetic
field, but becomes  positive
for $\q_0 R_9 > 1-  {R_{9} ^2\over 2{\a '} }  $.
In particular, type 0A theory with $\q _0 R_9\to 1 $  becomes type IIA string
theory,
and eq.~\uuu\ shows that the tachyon is absent.

The above formulas are valid in the perturbative regime,
 where the radius of the coordinate
playing the role of the eleventh dimension
is much smaller than the Planck length $l_P$.
The membrane interpretation \rbbb \ of the spectrum is
suitable for  extensions beyond perturbation theory.
Let us comment on how this  can be used to support
the  conjecture of  \BG\  that in the  flat-space
  type 0 theory,  the tachyon becomes massive at
 strong coupling.
In  string theory in NS-NS Melvin
background with   $\q R =1 $, the negative term in
the mass formula comes
from a normal ordering constant (see section 2.1 and \rios).
Let us assume that the same should be true
in the supermembrane theory.
In the case of  fermions
obeying antiperiodic boundary conditions on $x_{11}$,
the normal ordering constant in membrane theory
can  be computed
in the large radius limit, $R_9,R_{11}\gg l_P$ \russo .
Then,  instead of \rbbb , one
obtains ($w=1$):
\eqn\membri{
M^2=(4\pi^2  R_9R_{11}T_2 ) ^2 -   8 T_2 R_{11}^2 R_9^2
\sum_{(n,n')\neq (0,0)} [1- (-1)^n] (n^2R_{11}^2+{n '}^2 R^2_{9}
)^{-{3\over 2}}\ .
}
The infinite sum can be expanded in the region
 $R_{11}\gg R_{9}$,
where $M^2$ reduces to \rios  , and in the
region $R_{11}\ll R_9 $, where one gets
\eqn\meb{
M^2= (4\pi^2  R_9R_{11}T_2 ) ^2 -  28
    \zeta (3) T_2 { R_{9}^2\over R_{11} }\big[
    1+O({R_{11}^2\over R_9^2})\big] \
,
}
or, in terms of type 0A parameters \pram ,
\eqn\mebb{
M^2\cong   {R_{9} ^2\over {\a '} ^2}\bigg(1-    {7   \zeta (3)\over \pi ^2} \
{1\over
g_{s0}^2}\bigg)\ .
}
This expression (which applies for  $R_9^2\gg \a' g_{s0}^2$)
 suggests,
  that instead of the tachyon at weak coupling
 we indeed get  a  state with $M^2 > 0$
   for $g_{s0}^2>  {7   \zeta (3)\over \pi ^2}$.

\subsec{\bf Comment on  M-theory at finite temperature}

M-theory at finite temperature $T$ may be studied  by considering
 the $d=11$  theory
in euclidean target space with periodic time $x_0$,
 and periodic coordinate $x_{11}$, and  with fermions antiperiodic in
 $x_0$.
This picture was recently used in \russo\ to compute
  the one-loop
 free energy in the supergravity limit,
and to determine
a critical temperature $T=T(g_s)$ which
interpolates between  the Hagedorn temperature  in the $d=10$ limit
$g_s\ll 1$ and $T\sim l_P\inv $ in
the $d=11$  limit $g_s\gg 1$.

Since the boundary conditions are similar, we
 can  thus relate  M-theory at finite temperature  to  M-theory
compactifications based on the models (a) and (b),
with $x_0$ playing the role of  $x_9$ and
$T=(2\pi R_0)^{-1}$.
The $SL(2,Z)$ symmetry of the model (b) compactification,
 if maintained at the  full quantum level,
would imply that the exact free energy  in M-theory should have  the symmetry
\eqn\ssyy{
F(g_{\rm eff},A,l_P)=g_{\rm eff} F( g_{\rm eff}\inv ,A,l_P)\ ,
}
where
\eqn\ggff{
g_{\rm eff} ={R_{11}\over R_0}=2\pi \sqrt{\a '} T g_s\ ,\ \
\ \   \ \ \ A=R_0R_{11}\ .
}
In terms of string-theory  parameters $g_s, T, \a ' $ this gives
\eqn\ssyyy{
T^{-1} F(g_s,T,\a ')=\bar T^{-1} F( \bar g_s, \bar T , \bar \a ')\
,
}
where
\eqn\ggfff{
\bar T=(2\pi g_s\sqrt{\a '})\inv\ ,\ \ \ \
\bar\a'=\a ' {R_{11}\over R_0}=\a '(2\pi \sqrt{\a '} T g_s)\ ,\ \ \
\bar g_s =
{g_s\over  (2\pi T \sqrt{ \a' } g_s)^{ {3\over 2} }\ . }
}
Thus the duality relates the low and high temperature regimes,
 as well as  the weak and strong coupling regions.

While the   bosonic field
 contributions to the one-loop  $d=11$
  supergravity free energy must
  have this symmetry \russo, it is not obvious
  why it should  be present in the full theory with fermions.
   Our point here was to note that
the symmetry \ssyyy\ of the finite temperature
 M-theory free energy
 is related to the conjectured
 $SL(2,Z)$  symmetry of the type 0B string theory.

\newsec{Type 0A tachyon mass  shift  in  $d=10$
R-R  magnetic flux background   }

\def \P {\phi}

\def \ma {  m^2_{\rm eff} }


Starting  with   M-theory in the  background \maaa\
 with antiperiodic fermions in $x_{11}$,
  corresponding  to type 0A theory in the background of the
 R-R  magnetic flux 7-brane \mccc,
 one has   the opposite
situation  to the one in the case of  type IIA theory
in the same R-R background   \CG:
while  $\q_0 R_{11}=0$  represents the
 flat  type 0A theory,
 $\q_0 R_{11} =1$ should
give a theory with periodic fermions in the eleventh dimension,
i.e.  type IIA  theory in flat space.
This implies
 that
 type 0A string theory in the   R-R
 7-brane background \mccc,\mzzz\  must become
 tachyon-free (i.e. the type 0A tachyon should
 get positive $M^2$) for a sufficiently large R-R magnetic field.

By using the perturbative correspondence between  type 0A theory
and type IIA  theory (see sections~1 and 2),
and interpolating between them
with the NS-NS
magnetic background \fff ,
 we  argued in section  4.2 that the  tachyon of type 0A
 theory (considered as a limit of
 9-d interpolating theory)
 should disappear at some magnetic
coupling
$\q_0 $ (cf. \uuu ).
If we start with the $d=10$   type 0A theory and  turn on
the R-R Melvin  background \mccc , \mzzz ,
the
tachyon should
disappear at some magnetic field $0< b_0R_{11} \leq 1$,
 since at $b_0R_{11} =1$
the theory is expected to describe the type IIA theory
which does not have a tachyon.

 As we have seen   in section 4, in  the M-theory
 interpretation
 of type 0 theory based on \maaa, the tachyon state may be
 identified with a particular winding  membrane
 (``twisted sector") state.
 That means, in particular, that the tachyon mass
  formula should contain a gyromagnetic  interaction term
  linear in R-R magnetic field, which is  responsible
   for changing the sign of $M^2$.
{}From the point of view of perturbative type 0 theory,
the tachyon is neutral with respect to the R-R field,
so such linear coupling must have a non-perturbative origin.\foot{This is
in contrast to  the 9-d reduction case,
where momentum and winding
 tachyon modes carry charges with respect to the  NS-NS
vector fields (see eq.~\mmfor\ and  below).
This is related to the following question: if the
type 0 tachyon may indeed be interpreted as a winding mode of
M-theory orbifold, then it should be complex, not real as it is in
perturbative type 0 theory. A natural conjecture then is that
the tachyon should  get a scalar partner and thus become
charged under the
R-R vector field in non-perturbative regime.}

One may still wonder if it is possible to find  some indication
 of the effect of shifting of the tachyon mass  by the R-R flux
  already at the level of
 perturbative type 0A theory.
In this section  we
shall consider the equation for the type 0A tachyon
in the  R-R Melvin background \mccc,\mzzz\
 and show   that,  as in
 \refs{\KT,\KTt}, the R-R flux shifts the value of tachyon
 (mass)$^2$ by a positive term, {\it quadratic} in the field.
 This gives a complementary argument in
 support of   the  conjecture \CG\
 that the  tachyon  should disappear at
 sufficiently strong  magnetic R-R field.

Our starting point will be the effective action of type 0A theory
for the tachyon $T$ from (NS--,NS--) sector
 and massless fields:
the  (NS+,NS+) fields $(G_{mn},B_{mn},\P)$,
the R-R fields $(A_m,A_{mnk})$ and their
`doubles'  $(\bar A_m,\bar A_{mnk})$
from the twisted sector  of type IIB orbifold.
All type 0  tree amplitudes which involve only the fields
 from the (NS+,NS+)
and (R+,R+) (or (NS+,NS+) and (R--,R--))
sectors are { identical}  to those in the type II theory --
 in spite of the absence of
 space-time supersymmetry, the world-sheet supersymmetry implies
the same restrictions on the tree-level string effective action
as in the  type II theory (e.g., the absence  of  $\a'$ corrections
to 3-point functions) \KT.
 The leading
 second-derivative terms for the  NS-NS fields $(G_{mn},B_{mn},\phi)$
fields  and $T$ thus have the standard form
\eqn\grvi{
S_1 = -2 \int d^D x \sqrt G e^{-2\phi} \bigg[
 R + 4 (\del \phi)^2  - {\textstyle {1 \ov 12}}  H^2_{mnk} -
 \four  ( \del T)^2  - \four
 m^2_0 T^2   \bigg] \ ,
  }
  where $m^2_0= - {2\ov \a'}$ is the
  tachyon mass.\foot{The world-sheet supersymmetry also
constrains
the tachyonic sector of the action:
 all NS-NS amplitudes involving {\it odd} powers of
$T$ (e.g.,  $T^3$)
vanish. One is able to show also  that  there is no $ T^2 H^2_{mnk} $
term in the action \TP. For a recent
discussion of type 0 actions and their T-duality properties see
\ort.}
 The  leading  R-R terms in the action of  type 0A theory
are \KT\
\eqn\sta{
S_2 =  \int d^D x \sqrt G \ \bigg[   (1 + T^2) (
 |F_2|^2     +   |\bar F_2|^2
+   | F_4|^2   +    |\bar F_4|^2 )
+ \  | F_2 \bar F_2| T \ + | F_4 \bar F_4| T \bigg]   \ , }
where
$ | F_{n} \bar F_{ n}| \equiv { 1 \ov n!}
F^{m_1...m_n}\bar F_{m_1...m_n}$, \ $F_n = d A_n$.
This action  may be diagonalized by introducing
the ``electric" and ``magnetic"
combinations $
F_{n\pm} = { 1 \ov \sqrt 2} (F_n \pm \bar F_n)$.

It is possible to embed   any solution of type II theory into
 type 0 theory  by  setting the twisted sector fields to zero:
$\bar F_n =0, \ T=0$
 (for $T=0$ there will be no  non-trivial mixing
between the R-R forms from the standard and ``twisted" sectors)
\refs{\KT,\KTt}.
In particular,
  the R-R 7-brane background \mccc , \mzzz\  is
  also a solution ($F_{mn} \not=0, \bar F_{mn}=0$) of  type 0A string theory.

 The  equation for the  type 0A tachyon fluctuation
  in the  type 0A counterpart of
 type IIA
   R-R  vector background
    follows
 from the action \grvi,\sta\
\eqn\actio{
S=-2 \int d^{10}x \sqrt{-G} \bigg( e^{-2\phi}\big[R+4 (\del\phi)^2
- \four (\del T)^2  -  \four  m^2_0 T^2
\big]
-\four F_{mn }F^{mn} (1 + \ha T^2 )  \bigg) \ .
}
It follows from here  that  (part of) the
  effect  that switching on   R-R magnetic flux
shifts the tachyon (mass)$^2$ in  positive direction
appears  to be   visible  already  in
the weakly coupled type 0A  theory:
  the tachyon mass is indeed  shifted by a term quadratic
in the  magnetic R-R  flux
\eqn\taac{
- {1 \ov \sqrt G e^{-2 \phi} } \del_m ( \sqrt G
e^{-2 \phi} G^{mn} \del_n T)
 + \ma   T  =0 \ , }
\eqn\mas{
 \ma  = m^2_0   + \ha F^2_{mn} e^{2 \phi}     \ .
}
This  perturbative equation
may  not be, of course,  sufficient  to argue
that the spectrum of the scalar $T$ fluctuations
  in the
R-R Melvin background \mccc , \mzzz \
 becomes non-tachyonic  for sufficiently large field.
As follows from \mzzz,
 the dilaton (effective string
coupling)  blows  up  at large $r$, and
thus the above tree-level
$d=10$ effective action  is not applicable there.

We  may try, however, to   concentrate at
the near-core (small $r$)
region where the weakly-coupled type
0 string description should be  valid.\foot{Since we are  going
 to consider $r\to 0$, i.e.,
  in particular,  $r\sim \sqrt{\a'}$, we should
  be prepared to include all $\a'$ corrections.}
The maximum of $\ma (r) $  in \mas\ computed using \mccc,\mzzz\
(here  $\q_0$  denotes the value of $\q$ in type 0 theory) 
\eqn\bma{
 \ma  = m_0^2   +     { 4 \q ^2_0  e^{2 \phi_0}  \ov (1 + \q ^2_0 r^2 )^{5/2}
}\  }
  is reached at $r=0$
  \eqn\ama{
 \ma (0)   = -{  2\ov  \a'}  +     { 4 \q ^2_0  g^2_{s0 }
   }   \ , \ \ \ \ \ \ \ \
\
g_{s0} = e^{ \phi_0}
 \ .}
Assuming  that the string coupling
$g_{s0}= {  R_{11} \ov \sqrt{\a'} }$
is  small,
we would  conclude  from \ama\ that
the  critical value of the  magnetic field parameter at which
the tachyon becomes effectively  massless near  $r=0$ is
$
 \sqrt {2 \a'} \q_*  g_{s0} =  1   . $
It is clear that   $\a'$ corrections may  become
important here: since we are assuming $ g_{s0} < 1$,
  we get  $\a' \q^2_* > 1$.\foot{Indeed,
the scalar curvature
$
{R(r)}=- {3\q ^2 (4+3 \q ^2_0 r^2)\over (1+\q ^2_0 r^2 )^{5/2} } ,
$
which reaches its  maximum at $r=0$ (${ R}(0)=-12 \q ^2$),
is then  strong in string units:  $\a' { R} >  1 $.}

Still,  it is possible   that $\a'$ corrections do not
qualitatively
change    the  structure of the expression for the mass
 \bma,\ama\ so that this effective field theory result is indeed
 an indication that the tachyon  mass is shifted
 by  strong magnetic R-R field.\foot{A similar effect of
  shifting of  the type 0B
tachyon mass by the  R-R 5-form flux
 was observed in \refs{ \KT,\KTt}. There the dilaton was constant, so
 the theory was weakly coupled everywhere in space, and
 it was found that  one is to require that  the scale
 of the $AdS_5 \times S^5$ solution, i.e.
$L= ( 4 \pi g_s N)^{1/4} \sqrt{ \a' }$,  should be  of order
$ \sqrt{ \a' } $, i.e. sufficiently small, to
change the sign of the tachyon mass squared.}
As already mentioned at the beginning of this section,
non-perturbative corrections are expected to modify
\ama\ further, inducing a term  linear  in the R-R
 magnetic field.

\def\p{\phi }

\newsec{\bf A  class of magnetic flux-brane solutions in $d=11$}
By applying T and S duality transformations to the ``seven-brane''
solution
\dosa
, \aave\ one can generate other similar  flux brane backgrounds.
For example,
T-dualities along the $x_s$ directions give
 ``flux p-brane'' (or Fp-brane)
solutions with $p <6$.
The case of $p=4$  or  F4-brane solution is
  $$
ds^2_{10A}= f^{1/2}
\big[-dt^2+dx_1^2+dx_2^2+ dx_3^2+dx_4^2 + { f}\inv
(dx_5^2+ dx_6^2) +
{\tilde f\inv} dx_9^2+  dr^2+
{r^2 f\inv \tilde f\inv }  d\varphi ^2\big]\ ,\
$$
\eqn\ssff{ e^{2(\phi -\phi_0)}= f^{{1\over 2}} \tilde f\inv \ ,\ \ \
A_{\varphi 56 }=\q r^2 f\inv
\ ,\ \ \ \ B_{\vp  9}=\b  r^2\tilde f\inv \ .
}
This is the dimensional reduction along $x_{11}$ of the following
$d=11$  solution
$$
ds^2_{11}= (f\tilde f )^{1/3}
\big[-dt^2+dx_1^2+dx_2^2+ dx_3^2+dx_4^2 $$ \eqn\ryyy{  + \ { f\inv}
(dx_5^2+ dx_6^2)+{ \tilde f\inv } (dx_9^2 +d\x ^2) +dr^2+
{r^2 f\inv \tilde f\inv }  d\varphi ^2\big]\ ,\
}
\eqn\rry{
C_{\varphi 56 }=\q r^2 f\inv
\ ,\ \ \ \ \ \ \ \ \ C_{\vp  9 11  }=\b r^2 \tilde f\inv \ .
}
Notice  a remarkable   symmetry of  this background
under the interchange of $f$ and $\tilde f$ or
   $\q  $ and $\b  $, combined with relabelling of  the
   coordinates
   $(5,6) \leftrightarrow (9,11)$.
   This is another   11-d counterpart     (in addition to
   \surr,\ccf)  of
   our basic 2-parameter  magnetic background \melv,\meel.

In fact, this  is a particular case of  a
more general solution with four independent
magnetic field parameters\foot{Note that  this  metric
  has a structure similar
to that of the background describing  supersymmetric  \PT\
intersection of four  M5-branes \tset.
Dimensional reduction of this background to $d=4$
gives  a 4-parameter generalization of the Melvin
solution  constructed in \EMP,
where  also a more general class of  composite
$d=4$
black hole in magnetic field solutions was found
(these are somewhat different from
$d=4$ reductions of backgrounds   discussed below in section 7).}
$$
ds^2_{11}= (f_1 f_2 f_3 f_4)^{1/3}
\big[-dt^2+{ f_1\inv } (dx_1^2+dx_2^2)+{ f_2\inv }( dx_3^2+dx_4^2 )
$$ \eqn\rryy{
+\ { f_3\inv }
(dx_5^2+ dx_6^2)+{ f_4\inv } (dx_7^2 +dx_8^2) +dr^2+
{r^2 f_1\inv f_2\inv f_3\inv f_4\inv}  d\varphi ^2\big]\ ,\
}
$$
C_{\varphi 12 }=\q_1r^2 f_1\inv
\ ,\ \ \ \  \ \ C_{\vp  34 }=\q_2 r^2 f_2\inv\ ,\ \
$$
\eqn\uio{
C_{\vp  56 }=\q_3 r^2 f_3\inv \ ,\ \ \ \ \ \ \ \  C_{\vp  78 }=
\q_4 r^2 f_4\inv \ , \ \ \ \ \ \ \ \ \  f_i=1+\q_i^2 r^2\ .
}
This solution may be obtained by starting with
 \ryyy,\rry\ and mixing $\vp$
 with other
 ``parallel" coordinates.
To explain  this construction
 in more detail let  us first rederive  \ryyy,\rry.
 Let us
 start with  flat $d=11$  Minkowski space  with $C_3=0$,
and make a formal coordinate shift $\vp \to \vp + \q_1 x_8$
as in \maaa.
 By dimensional reduction in $x_8$ and  T-duality
in $x_1,x_2$,
we get a type IIA solution with  the R-R 3-form having the
 $A_{\vp 12}$ component.
Uplifting it  back  to eleven dimensions, this
gives the background \rryy\ with $\q_2=\q_3=\q_4=0$,
i.e. \ryyy\ with $\b=0$ or \surre.
Next, we start with this non-trivial
$d=11$  background and  repeat the same procedure:
replace  $\vp \to \vp +\q_2 x_2$, and make dimensional reduction in $x_2$.
We obtain a type IIA solution with
$A_\vp $ and $B_{\vp 1}$ fields.
 By T-duality in $x_3,x_4$, and uplift,
  we get \ryyy, or  \rryy\ with $\q_3=\q_4=0$.
Iterating this procedure, we introduce $\q_3$ and $\q_4$
and finally  get \rryy,\uio.
This background  can be further generalized by
 making a  shift $ \vp \to \vp  + b_i' x_i$,
 leading to magnetic solutions
 in 10 dimensions with more free  magnetic parameters.

Finally, let  us mention  another example   of  $d=11$
 solution  which
 is  obtained by  uplifting  to 11 dimensions
  the electric/magnetic F2-brane
$$
ds^2_{11}= \tilde f^{1/3} f^{2/3}
\big[-dt^2+dx_1^2+dx_2^2+ { f\inv }( dx_3^2+dx_4^2 +
dx_5^2+ dx_6^2)
$$
\eqn\zzy{
+\  \tilde f\inv  dx^2_{9} + { f\inv \tilde f\inv }dx^2_{11} +dr^2+
{r^2 f\inv \tilde f\inv }  d\varphi ^2\big]\ ,\
}
$$
C_{t12 }=\q r^2 f\inv
\ ,\ \ \ \ \ \ \  C_{\vp  9 11 }=\b  r^2\tilde f\inv \ .
$$
Interchanging $t$ and, say,  $x_3$, gives
another magnetic solution.

All such  $d=11$ magnetic flux brane solutions
 are non-supersymmetric and (like the background  \surre,\ccf )   may be
 perturbatively  unstable in certain region of parameter space.


\newsec{Superpositions of  magnetic flux  branes    and D-branes}


So far  we described
some   brane-type  solutions with magnetic fluxes:
instead of magnetic fluxes through  spheres
 (which describe magnetic charges), they have axially symmetric
magnetic fluxes  through planes.
 Other classes of flux-brane
 solutions  were presented in \refs{\GS,\suf,\gal}.

It is of some interest to investigate ``hybrid"
solutions which have both brane charges and
Melvin-type  magnetic fluxes.
One  way  to construct them is to start with M5-brane or M2-brane solution
in
$d=11$,
pick up a 2-plane (in parallel or transverse directions)
with the metric $ dx^2_1 + dx^2_2 = dr^2 + r^2 d\vp^2$, and
then shift
$\vp \to \vp + \q x_{11}$. Dimensional reduction  along $x_{11}$
(combined with U-dualities) then generates
a superposition  of a
D-brane with the  K-K  Melvin  magnetic flux through  the
$(x_1,x_2)$ plane.

The  twist $\vp \to \vp + \q \x $ breaks supersymmetry of the original
M-brane
background.   As in the case of the absence
 of M-brane charges,   the magnetic
parameter $\q $
allows one to    smoothly
 interpolate between periodic and antiperiodic boundary conditions
 for the fermions.  Since  all  solutions of type II string
  theory  may  be embedded
 into  type 0 string theory (see section 5),
 and in view of the (perturbative and non-perturbative)
 relations between type II and type 0
 theories discussed above,
 this magnetic $\q $-interpolation
  may allow one
 to relate  D-branes in type II theory  to D-branes in type 0
 theory.

For example, consider
a type IIA  Dp-brane superposed with an axially symmetric  magnetic flux
with
parameter $\q$.
While for  small string coupling  or small $\q R_{11}$
the  $d=10$ type IIA  description  is applicable, this is no longer so for
 $\q R_{11} \sim 1$ (the effective 11-d radius becomes  large),
 i.e. one can no longer  use the supergravity approximation
  based on the dimensionally reduced
($d=10$)
solution.
The description of the  $\q R_{11}\sim 1$ case should be based
on  M-theory with  fermions being {antiperiodic}   along $x_{11}$.
Following \refs{ \CG ,\GS },  one may then
conjecture  that the result should be
equivalent
to a  Dp-brane in type 0 theory
in a R-R Melvin background
with a small magnetic field  with  $\q_{0} R_{11}=1- \q R_{11}$.

Below  we   shall  present examples of  such ``hybrid" solutions
with a hope that they may be useful in future   studies
of   connections  between D-branes of
type II  and  type 0  theories.

\def \r {\rho}

Starting with a $d=11$    M-brane solution
 and mixing the angular  coordinate
$\vp$ in the  $(x_1,x_2)$ plane  with  $x_{11}$ to add magnetic flux,
 one has several options:
$x_1,x_2,x_{11}$   may be  parallel directions to M-brane;
$x_1,x_2$  may be  parallel  while  $x_{11} $  may be  transverse;
 $x_{11}$ may be parallel while  $x_1,x_2$ may be transverse, etc.
There are also several options of
making  dimensional reduction and  applying
dualities  to generate  various  $d=10$ ``hybrid"
  brane  backgrounds.
Let us list  some of the  solutions that can be obtained in this way.
We will use the notation
$
H_n=1+{L^n\over \r^n}\
$ for the harmonic function of an M-brane
($\rho$ is the  radial direction transverse to the brane).

{}From  the   M5-brane  background with a twist in  the parallel
 2-plane in the metric
\eqn\bfive{
ds^2_{11}= H_3^{-1/3} [ -dt^2+dx_3^2 +dx_4^2+dx^2_{11}
 + dr ^2 + r^2 (d\varphi + \q
d\x )^2 ] +
H_3^{2/3} [ d\r^2 + \r^2 d\Omega_4^2 ]\ ,
}
one obtains the D4-brane with a R-R magnetic flux on the  world-volume:
\eqn\dfour{
ds^2_{10A}=f^{1/2}  \bigg(
H^{-1/2}_3\big[-dt^2+dx_3^2+dx_4^2+dr ^2+
{r^2 f\inv } d\varphi^2\big]+H^{1/2}_3\big[   d\r^2 + \r^2 d\Omega_4^2
\big]\bigg)\ ,
}
\eqn\bdil{
e^{2\phi}= H_3^{-1/2}(\r)  f^{3/2}(r) \ ,\  \ \ \ \
\ A_\varphi =\q r^2 f\inv   \ ,\ \ \ \ \
 \ f(r )=1+\q ^2r ^2\ \ .
}
 T-duality then  gives a  (smeared version of)
  D3-brane solution with  R-R 2-form flux,
\eqn\dfff{
ds^2_{10B}=f^{1/2} \bigg( H_3^{-1/2}\big[-dt^2+dx_3^2+dr^2+
{r^2 f\inv } d\varphi^2\big]+H_3^{1/2}\big[  f\inv {dx_4^2}+ d\r^2
+ \r^2 d\Omega_4^2 \big]\bigg)\ ,
}
$$
e^{2\phi }=f(r)\ ,\ \ \ \ \  A_{\varphi 4}= \q r^2 f\inv(r) \ .
$$
Starting with  the M5-brane  with the ``parallel" twist  \bfive\
 smeared along $x_{11}$
\eqn\xfive{
ds^2_{11}=H_2^{-1/3}\big[-dt^2+dx_1^2+dx_2^2+dx_3^2+dr^2 + r^2
(d\vp + \q  dx_{11} )^2 \big]+H_2^{2/3}
\big[ dx^2_{11} +d\r^2 +\r^2 d\Omega_3 ^2\big]\ ,
}
one obtains the NS5-brane with R-R Melvin magnetic flux:
\eqn\zztwo{
ds^2_{10A}=\hat f^{1/2}\bigg(-dt^2+dx_3^2+dx^2_4 +dx_5^2+dr^2 +
{r ^2 \hat f\inv  } d\varphi^2 + H_2\big[
 d\r^2+\r^2 d\Omega_3^2 \big]\bigg)\ ,
}
$$
e^{2\phi }= H_2 \hat f^{3/2}\ ,\ \ \ \
 A_\vp ={\q r^2 H_2\inv  \hat f\inv }
\ ,\ \ \ \ \ \hat f(r,\r)\equiv 1+{\q ^2r^2\over H_2 (\r) }\ , \ \ \
H_2 = 1 + { L^2\ov \r^2} \ .
$$
The  smeared M5-brane with two extra  isometries  with transverse
twist
\eqn\xfive{
ds^2_{11}=H_1^{-1/3}\big[-dt^2+dx_3^2+...+dx_6^2+dx^2_{11} \big]+H_1^{2/3}
\big[ dr^2 + r^2 (d\vp + \q  dx_{11})^2 +d\r^2 +\r^2 d\Omega_2 ^2\big]\ ,
}
gives the  D4-brane with R-R magnetic flux   in  transverse space
\eqn\dftwo{
ds^2_{10A}= \hat f^{1/2}\bigg(   H_1^{-1/2}
\big[-dt^2+dx_3^2+...+dx_6^2\big]+
H_1^{1/2}  \big[  dr ^2+
{r ^2  \hat f \inv } d\varphi^2 + d\r^2+\r^2 d\Omega_2^2 \big]\bigg)\ ,
}
$$
e^{2\phi }= H_1^{-1/2} \hat f^{3/2}\ ,\ \ \ \ A_\vp =\q r^2 H_1 \hat f\inv
\ ,\ \ \ \ \ \hat f(r,\rho  )\equiv 1+\q ^2 r^2 H_1(\r)\ .
$$
{}From the M2-brane with three extra isometries, one obtains
a D2-brane solution with transverse R-R magnetic flux:
\eqn\ddtwo{
ds^2_{10A}=f^{1/2}\bigg( H_3^{-1/2}\big[-dt^2+dx_3^2+dx_4^2\big]+
H_3^{1/2}\big[  dr ^2+  {r ^2  f\inv } d\varphi^2 + d\r ^2+\r ^2
d\Omega^2_4\big]\bigg)\ ,
}
$$
e^{2\phi } = H^{1/2}_3 f^{3/2}
 \ ,\ \ \ \ \ \ \    \ A_\varphi =\q r^2 f\inv\ .
$$
One may apply a similar procedure of a twist and dimensional reduction to
the 11-d plane wave background, getting a superposition of D0-brane and R-R
flux 7-brane. Here the R-R  1-form will have both electric and magnetic
components.

Another way to construct ``hybrid" solutions is
to start with  conformal NS-NS sigma model describing NS5-brane
 and to replace the 5 free   parallel directions  by another conformal
 theory
--
 the 2-parameter magnetic model \lagg.
 The resulting background  has the following  metric and dilaton  \prt
\eqn\dggg{
ds^2_{10A}=-dt^2+dx_1^2+dx_2^2+dr^2+ {\tilde f\inv }
\big[ r^2 \big( d\varphi+\q  dx_9 \big )^2+ dx_9^2 \big]+H_2\big(
   d\r^2 + \r^2 d\Omega_3^2 \big)\ ,
}
\eqn\yop{
e^{2\phi }= e^{2\p_0} H_2\tilde f\inv\ , \ \ \ \ \  \  \tilde f=
 1+\b ^2r^2 \ , \ \ \ \   H_2 = 1 + {L^2 \ov \r^2} \ ,  }
 as well as NS-NS  2-form which is a combination the one in \meel\
 with NS5-brane one, i.e.
 \eqn\gii{ dB_2 = { 2 \b r \ov ( 1+\b ^2r^2)^2} dr \wedge d \vp \wedge
 dx_9
 + L^2 d {\rm Vol}(S^3) \ . }


\bigskip


{\bf Acknowledgements}

\noindent
We are grateful to S. Frolov
 for very  useful discussions.
The work of A.A.T. is partially supported by the DOE grant
  DE-FG02-91ER40690,
   PPARC SPG grant 00613,  INTAS
  project 991590 and
 CRDF Award RPI-2108.
 Part of this work was done while
A.A.T. was   participating in the
M-theory program at ITP, Santa Barbara
supported by the NSF grant PHY99-07949.

\appendix{A}{ Constant magnetic field background}


For completeness, let us mention also another example of
a  magnetic background  which
 in a sense  comes  closer than Melvin solution
to the aim of modelling
 a uniform  magnetic field in  closed string theory.

In general relativity, magnetic flux lines gravitationally attract
 each other and thus  tend to
concentrate near $r=0$. A way to have a homogeneous
 (covariantly constant)
magnetic field is to add a rotation:
 intuitively, it is clear that
 magnetic flux lines which would
otherwise cluster near $r=0$ would
 then repel due to  centrifugal forces.
Such NS-NS  background  was discussed in \refs{\constant ,\heterotic }
by giving a magnetic Kaluza-Klein interpretation to  the following
 gravitational wave background
(related to a non-semisimple WZW model  \NW)
\eqn\nappi{
ds^2_{10A}= -dt^2+dx_s^2+dx_9 ^2+ 2  \bb r^2 d\varphi (dx_9 - dt )
 +dr^2+r^2d\varphi ^2\ ,
}
\eqn\bee{
B_2 = \bb r^2 \ d\varphi \wedge (dt - d\y ) \ , \ \ \ \ \ \ \  \ \
\bb
=\const  \ .
}
As was shown in \refs{\constant, \heterotic}
the corresponding conformal string model
is exactly solvable in terms of free
fields and has Hamiltonian similar  to \hail.
A more general 3-parameter class of solvable models
interpolating between \lagg\ and \nappi\ was discussed
in \exactly\ (see also \tcqg).

Applying U-duality transformations to \nappi,\bee\ one can
construct related
supergravity solutions with R-R magnetic fields.
One way to do that is first to lift  the above background to 11
dimensions and
then
reduce down along a different
spatial direction.
The  $d=11$  solution corresponding to \nappi,\bee\ is
\eqn\once{
ds^2_{11}= -dt^2+dx_s^2   +d\y ^2+ 2\bb r^2 d\varphi (dx_9 - dt)
+dr^2+r^2d\varphi ^2+dx_{11}^2\ ,
}
\eqn\mon{
C_3 =\bb r^2\ dx_{11}\wedge d\varphi \wedge (dt - d\y)\ . }
Dimensional reduction in the $\y$ direction then gives ($x\equiv
x_{11}$)
\eqn\con{
ds^2_{10A}= -(dt+ \bb r^2 d\varphi )^2 +dr^2+r^2d\varphi ^2+dx_s^2
+dx^2\ ,
}
\eqn\ter{
B_2 =\bb  r^2 \ d\varphi \wedge dx  \ ,\ \ \ \ A_1 =\bb r^2 d\vp \
,\ \ \ \
A_3 =\bb r^2  d\varphi \wedge dt \wedge  dx \ . }
A special property of these  backgrounds
is  that the gauge field strength and
 curvature invariants  are {\it constant}.

The NS-NS model  \nappi\ has tachyonic modes with high spin
for arbitrarily  small
values of the magnetic field \heterotic .
It would be interesting to study the stability of the R-R model
\con,\ter .
 \listrefs
\vfill\eject
 \end